\documentclass[epj-spec]{svjour}
\usepackage{graphics}
\usepackage{epsfig}
\usepackage{amsmath}
\usepackage{amssymb}
\usepackage{epsf}
\usepackage{graphicx}
\usepackage{cite}
\usepackage{ulem}
\usepackage{rotating}
\usepackage{slashed}

\newcommand{\sqrtsnn}{\sqrt{s_{_{{\rm N N}}}}}
\newcommand{\sqrts}{\sqrt{s}}
\newcommand{\A}{{\rm A}}
\newcommand{\dd}{{\rm d}}
\newcommand{\lqcd}{\Lambda_{_{{\rm QCD}}}}
\newcommand{\msbar}{\overline{\rm MS}}
\newcommand{\kt}{k_\perp}
\def\qhat{\hat{q}}
\def\cO#1{{{\cal{O}}}\left(#1\right)}
\def\pt{p_{_\perp}}
\def\X{{\rm X}}
\def\epem{e^+e^-}
\def\alphas{\alpha_s}
\def\alphabar{{\overline{\alpha}_s}}
\def\nch{n_{\rm ch}}
\def\mf{\mu}
\def\mh{m_h}
\def\xbj{x_{_{\rm Bj}}}
\def\meaneps{\langle\epsilon\rangle}
\def\ptpi{p_{_{\perp_h}}} \def\ptgamma{p_{_{\perp_\gamma}}}
\def\gampi{\gamma$--$h\ } 
\def\z{z_{_{\gamma h}}}
\def\picut{p_{_{\perp_\pi}}^{\rm cut}}
\def\gacut{p_{_{\perp_\gamma}}^{\rm cut}}
\def\url#1{\href{#1}{{\tt #1}}}

\begin{document}
\title{(medium-modified) Fragmentation Functions}
\author{Fran\c{c}ois Arleo}
\institute{LAPTH\footnote{Laboratoire d'Annecy-le-Vieux de Physique Th\'eorique, UMR5108}, Universit\'e de Savoie, CNRS, BP 110, 74941 Annecy-le-Vieux cedex, France}
\date{Received: date / Revised version: date}
\abstract{
In this short review paper, we discuss some of the recent advances in the field of parton fragmentation processes into hadrons as well as their possible modifications in QCD media. Hadron production data in $\epem$, deep inelastic scattering and hadronic collisions are presented, together with global analyses of fragmentation functions into light and heavy hadrons and developments on parton fragmentation in perturbative QCD at small momentum fraction. Motivated by the recent RHIC data indicating a significant suppression of large-$\pt$ hadron production in heavy-ion collisions, several recent attempts to model medium-modified fragmentation, e.g. by solving ``medium'' evolution equations or through Monte Carlo studies, have been proposed and are discussed in detail. Finally we mention the possibility to extract medium-modified fragmentation functions using photon--hadron correlations.
}
\maketitle
\section{Introduction and summary}

The fragmentation process --~that is the transition from a highly virtual time-like parton, $Q^2\gg\lqcd^2$, into a collimated bunch of hadrons~-- is a subject of first importance in QCD.
Similarly to structure functions in deep-inelastic scattering, the momentum distribution of hadrons produced in the final state cannot be determined from perturbation theory. The reason is twofold:
\begin{itemize}
\item the collinear splitting of one parton into two is divergent;
\item hadronization is a soft, non-perturbative, process.
\end{itemize}
Because of this soft and collinear sensitivity, the {\it fragmentation function} (FF) of a given parton $i$ with virtuality $Q$ into hadrons $h$ carrying an energy fraction $x$, $D_i^h(x, Q^2)$, cannot be determined from first principles and should be extracted from experimental measurements. However, as long as $Q$ is large as compared to $\lqcd$, the collinear divergences can be systematically factorized to all orders~\cite{Collins:1981uw},
\begin{equation}\label{eq:eveq}
D_i^h(x, Q^2) = \sum_k \int_x^1 \dd{z}\ K_i^k(z, Q^2, Q_0^2)\times D_k^h(x/z, Q_0^2),
\end{equation}
where the evolution of the parton $i$ into the parton $k$, with virtuality $Q_0$ and energy fraction $z$, is described by the evolution kernel $K_i^k(z, Q^2, Q_0^2)$ and can be computed in perturbative QCD (pQCD). Therefore, even though the theoretical objects $D_i^h(x, Q^2)$ are strictly speaking infinite, their scaling violation is under perturbative control and has been successfully tested in hadron production in $\epem$ collisions, e.g. at LEP~\cite{Kniehl:2000hk}.

Based on these $\epem$ data, several parametrizations of fragmentation functions into light hadrons (BKK~\cite{Binnewies:1994ju},  BFGW~\cite{Bourhis:2000gs}, KKP~\cite{Kniehl:2000fe}, Kr(etzer)~\cite{Kretzer:2000yf}, AKK05~\cite{Albino:2005me}) and into photons (BFG~\cite{Bourhis:1997yu}) have been performed. Despite the large amount of $\epem$ data, those FF sets still suffered from rather large uncertainties, especially at large $x$ and/or small $Q^2$.
The situation has however considerably improved recently: experimentally, new measurements in deep inelastic scattering (DIS) and hadronic collisions at RHIC have been made available, while on the theoretical side two of the most recent fits, HKNS~\cite{Hirai:2007cx} and DSS~\cite{deFlorian:2007aj}, now include error analyses (work is in progress for AKK08~\cite{Albino:2008fy}), in analogy to what has been performed for parton distribution functions (see e.g.~\cite{Pumplin:2001ct,Stump:2001gu,Botje:1999dj}). In the heavy-quark sector, the impressive results from the B-factories experiments (BaBar, Belle, CLEO) are expected to bring in the near future significant constraints on the non-perturbative component of FF into heavy hadrons.

At small values of $x\ll1$, or equivalently large values of $\xi\equiv\ln(1/x)$, the Modified Leading Logarithmic Approximation (MLLA)~\cite{Dokshitzer:1991wu,Dokshitzer:1982fh,Azimov:1985by} proved very successful in describing a variety of jet observables, such as single inclusive spectra or multiplicity distributions~(for a review on MLLA phenomenology, see~\cite{Khoze:1996dn}). The success of this approximation, dealing with perturbative quark and gluon degrees of freedom, supports the idea of Local Parton Hadron Duality (LPHD)~\cite{Azimov:1984np} which basically assumes that the shape of parton spectra remains virtually unmodified at the hadronization stage (only an overall normalization factor, ${\cal K}=\cO{1}$, is necessary to accommodate the hadron data). In some cases though, typically whenever hadrons are produced at large angles with respect to the jet axis, the MLLA predictions fail to describe the experimental measurements. Interestingly, the inclusion of terms of order $\cO{\alphas}$ in the solution of the evolution equations, i.e. strictly beyond MLLA, appears to improve significantly the agreement with data~\cite{Arleo:2007wn}.

Fragmentation processes have also received a lot of attention in the context of high-energy heavy-ion collisions where a dense and hot partonic system may be produced. Indeed, unlike in $p$--$p$ collisions, the quarks and gluons produced with large transverse momenta in the initial nucleon-nucleon binary collisions in heavy-ion scattering (i.e. when the two incoming nuclei overlap) propagate through the strong colour field --~whose size and lifetime are $\cO{10\ {\rm fm}}$~-- produced from the soft ``underlying'' event. Therefore, the full fragmentation process of hard partons should be affected by the presence of the dense QCD medium, at least in principle. A typical example is parton energy loss caused by medium-induced multiple gluon emission, a process which has been widely studied over the past decade (see e.g.~\cite{Baier:2000mf} for reviews). This mechanism may be responsible for the spectacular ``jet quenching'' phenomenon reported by the PHENIX~\cite{Adcox:2001jp,Adler:2003qi} and STAR~\cite{Adler:2002xw,Adams:2003kv} experiments at RHIC, that is the suppression of single inclusive hadron spectra by a factor of 5 in central Au--Au collisions with respect to (properly scaled) $p$--$p$ scattering. A lot of effort is now devoted in understanding how the parton multiple scattering process may affect QCD evolution. Based on the extensive phenomenology developed for jets produced in the ``vacuum'', i.e. in $\epem$, DIS and hadronic collisions, several attempts to model fragmentation functions in heavy-ion collisions have been suggested, either by solving medium-modified evolution equations~\cite{Borghini:2005em,Armesto:2007dt,Domdey:2008gp,Quiroga:2008hp,Dremin:2006da} or through Monte-Carlo studies~\cite{Zapp:2008gi,Renk:2008pp,Cunqueiro:2008hp}.

In this short review paper, we will highlight the recent advances on fragmentation functions\footnote{The field of spin-dependent fragmentation functions and transversity is not addressed in the present paper. For a review on that subject, see Ref.~\cite{Barone:2001sp}.} as well as their possible modifications in QCD media (see also~\cite{Trento:2008wo} for a comprehensive overview of that subject). Section~\ref{sec:ffvac} is devoted to fragmentation processes in the vacuum. After discussing the new data available (Section~\ref{sec:datalight}), emphasis is then put on the recent FF sets and their improvements with respect to older studies (Section~\ref{sec:globalfits}), the developments beyond MLLA at small $x$~(Section~\ref{sec:mlla}) as well as parton fragmentation into heavy hadrons (Section~\ref{sec:heavyquark}). Moving to medium-modified fragmentation processes in Section~\ref{sec:ffmed}, data in electron--nucleus and nucleus--nucleus collisions are first summarized (Section~\ref{sec:datamedium}). After having briefly recalled the basics of parton energy loss processes (Section~\ref{sec:energyloss}), we discuss the modelling of QCD evolution in presence of a QCD medium (Sections~\ref{sec:rescaling} and~\ref{sec:evolutionmedium}) and review recent attempts to develop parton showers in nucleus--nucleus collision in Section~\ref{sec:partonshowers}. Finally, we show in Section~\ref{sec:correlations} how performing photon--hadron momentum correlations may help to extract medium-modified fragmentation functions in heavy-ion collisions.

\section{Fragmentation function studies}\label{sec:ffvac}

\subsection{Data}\label{sec:datalight}

\subsubsection{$\epem$ annihilation}\label{sec:epem}

The measurements of the {\it total} fragmentation function in $e^+e^-$ annihilation, or scaled-energy distributions,
\begin{equation}\label{eq:momdist}
F^h(x, s) = \frac{1}{\sigma_{\rm tot}}\ \frac{\dd\sigma}{\dd{x}}(e^+e^-\to{h}\X)
\end{equation}
of light hadrons were first performed at DORIS/PETRA at DESY (ARGUS~\cite{Albrecht:1989wdAlbrecht:1987ew}, CELLO~\cite{Behrend:1989ae}, JADE~\cite{Bartel:1981sw}, TASSO~\cite{Althoff:1982dhBrandelik:1980iy} experiments) at centre-of-mass energies $\sqrt{s}=12$--$36$~GeV, as well as at SLAC~PEP (HRS~\cite{Derrick:1985wd}, MARK~II~\cite{Schellman:1984yzdelaVaissiere:1984xg}, TPC~\cite{Aihara:1986mvAihara:1988su}) at $\sqrts=29$~GeV, and at KEK~TRISTAN (TOPAZ~\cite{Itoh:1994kb}) at $\sqrt{s}=58$~GeV. More recently, hadron production in $\epem$ collisions has also been measured at the $Z^0$ pole ($\sqrt{s}=91.2$~GeV) both at CERN LEP-I (ALEPH~\cite{Buskulic:1994ft,Buskulic:1995sw,Barate:1996fi,Barate:1998cp,Heister:2003aj}, DELPHI~\cite{Abreu:1998vq,Abdallah:2006ve}, OPAL~\cite{Akers:1994ez,Abbiendi:1999pi,Abbiendi:1999ry}) and at the SLAC linear collider (SLD~\cite{Abe:1998zs,Abe:2003iy}), and at higher energy, $\sqrt{s}=133$--$209$~GeV, at LEP-II (DELPHI~\cite{Abreu:1999vs,Abreu:2000gw}, L3~\cite{Achard:2004sv}, OPAL~\cite{Alexander:1996kh,Abbiendi:1999sx}).

\begin{figure}[htb]
    \begin{center}
      \includegraphics[width=9.6cm]{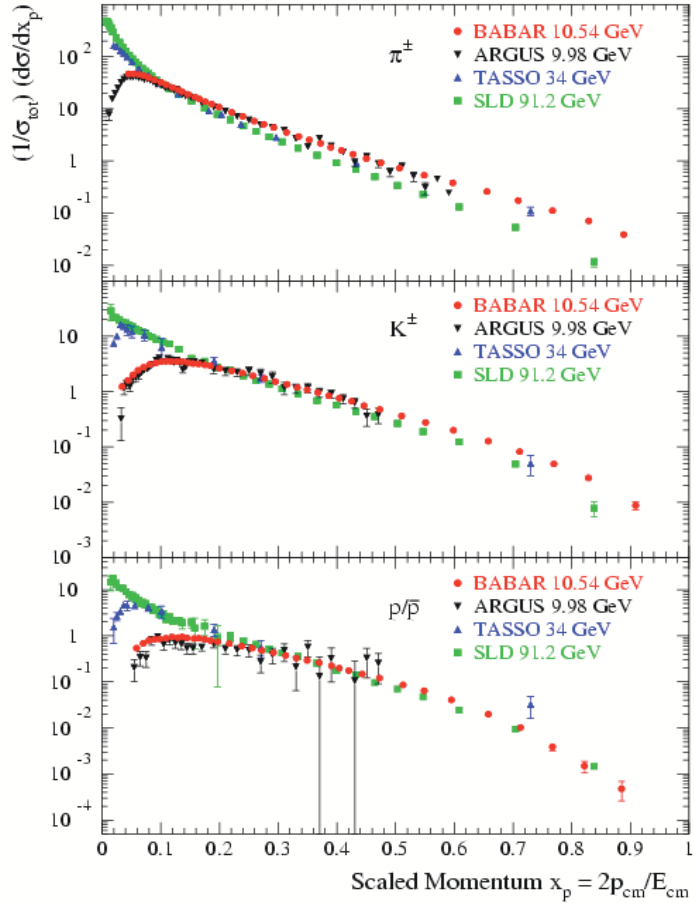}
    \end{center}
    \caption{BaBar preliminary measurements of scaled-energy distributions of  data $\pi^\pm$, $K^\pm$, and $p$/$\bar{p}$ in $\epem$ collisions at $\sqrt{s}=10.54$~GeV. Taken from Ref.~\cite{Anulli:2004nm}.}
    \label{fig:babar}
\end{figure}

The distributions have been measured for both unidentified and identified ($\pi^\pm$, $K^\pm$, $K_s^0$, $p$/$\bar{p}$, $\Lambda$/$\bar{\Lambda}$) hadrons on a wide range of momentum fractions, from $x\simeq5\ 10^{-3}$ ($\xi=-\ln x\simeq5$) at LEP/SLC up to $x\simeq 0.8$, the statistical uncertainty becoming however fairly large at high $x$. Hadron spectra inside gluon jets have been obtained from a selection of symmetric three-jet events, $\epem\to{q}\bar{q}g$ and using $b$-tagging techniques~\cite{Buskulic:1995sw,Barate:1998cp,Abdallah:2006ve,Abbiendi:2003gh,Abbiendi:2004pr}. Distributions in gluon jets are found to be softer than the ones from quarks jets (also reported are stronger scaling violations~\cite{Abbiendi:2004pr}) because of the larger colour charge of the gluon: the higher multiplicity making hadrons to be produced at smaller $x$ by momentum conservation. Moreover, several experiments (DELPHI~\cite{Abreu:1998vq}, OPAL~\cite{Abbiendi:1999ry}, SLD~\cite{Abe:2003iy}) were able to perform flavour-tagged measurements and to disentangle light-quark ($uds$) from $c$-quark and $b$-quark jets; note in particular that the OPAL experiment reported on hadron spectra in each {\it individual} light flavours~\cite{Abbiendi:1999ry}. A discrepancy was found between the NLO calculations using the FF parametrizations then available (BFGW, KKP, Kr) and the bottom and gluon jet data, despite a good agreement with flavour-inclusive and $udsc$-jet measurements~\cite{Abbiendi:2004pr}. For more detail about these data samples, we refer the reader to the global fit analyses~\cite{Bourhis:2000gs,Kretzer:2000yf,Albino:2008fy,Albino:2005me,Kniehl:2000fe,Binnewies:1994ju,Bourhis:1997yu,deFlorian:2007aj,Hirai:2007cx}~and the references therein.

Impressive preliminary measurements on hadron spectra in $\epem$ annihilation have also been reported very recently from B-factories (BaBar, BELLE, CLEO) around the $\Upsilon$(4S)-mass, $\sqrts=10.58$~GeV, and below ($\sqrt{s}\lesssim7$~GeV) through initial-state radiation studies, $\epem\to\gamma\ {h}\ \X$. While those experiments are naturally suited to study heavy-quark hadrons (discussed in Section~\ref{sec:heavyquark}), it is worth to mention that light-hadron spectra can also be measured with an accuracy better than what has been achieved so far. For the sake of an illustration, preliminary measurements of $\pi^\pm$, $K^\pm$, and $p$/$\bar{p}$ inclusive spectra by the BaBar collaboration~\cite{Anulli:2004nm} are plotted in Figure~\ref{fig:babar} together with older data from ARGUS, TASSO, and SLD.

\subsubsection{DIS}\label{sec:dis}

Inclusive hadron spectra have also been studied experimentally in DIS events, mostly in $e^-p$ collisions at HERA by H1~\cite{Aid:1995up,Aaron:2007ds}, HERMES~\cite{Hillenbrand:2005pc}, ZEUS~\cite{Derrick:1995ca,Chekanov:2008hy} experiments but also in neutrino scattering, $\nu_\mu{p}$ and $\overline{\nu}_\mu{p}$, by NOMAD~\cite{Altegoer:1998py}. The virtuality range probed at HERA, $10\lesssim Q\lesssim100$~GeV, coincides fairly well with $\epem$ data, while being somewhat smaller ($1\lesssim Q\lesssim10$~GeV) with NOMAD.

In order to ease the comparison with $\epem$ collisions, the DIS events are analyzed in the Breit frame for which the photon has a virtuality but no energy. In the parton model, the struck (anti)quark has longitudinal momentum $p_z=Q/2$ and $p_z=-Q/2$ before and after the collision with the virtual photon. The hemisphere containing particles with negative $p_z$ is called the {\it current} hemisphere and should be directly comparable to $\epem$ collisions in which a quark is produced with energy $\sqrt{s}/2$, while the target hemisphere ($p_z>0$) describes the proton remnants. 

Quite remarkably, the momentum distribution~(\ref{eq:momdist}) of inclusive charged hadrons has been measured recently by H1 for 9 different intervals in $x$ and for 6 bins in virtuality $Q$, allowing for a systematic study of scaling violations of fragmentation functions~\cite{Aaron:2007ds}. In all the $(x, Q^2)$-bins for which these data can be compared to $\epem$ results, a very good agreement is obtained supporting the universality of the fragmentation processes (see also Refs.~\cite{Albino:2006wz,Kniehl:2000hk} for universality checks from $\epem$ to DIS) expected from factorization theorems. A small difference between the two scattering systems occurs at low energy scale, coming most probably from next-to-leading order corrections present in DIS and absent in $\epem$ annihilation. Comparing data with theoretical expectations reveals that NLO QCD predictions, using either KKP, Kr, or AKK05 FF sets, are not able to describe properly the $Q$-dependence of H1 data,  the scaling violation being much stronger in the experiment~\cite{Aaron:2007ds}.

\begin{figure}[htb]
    \begin{center}
      \includegraphics[width=8.6cm]{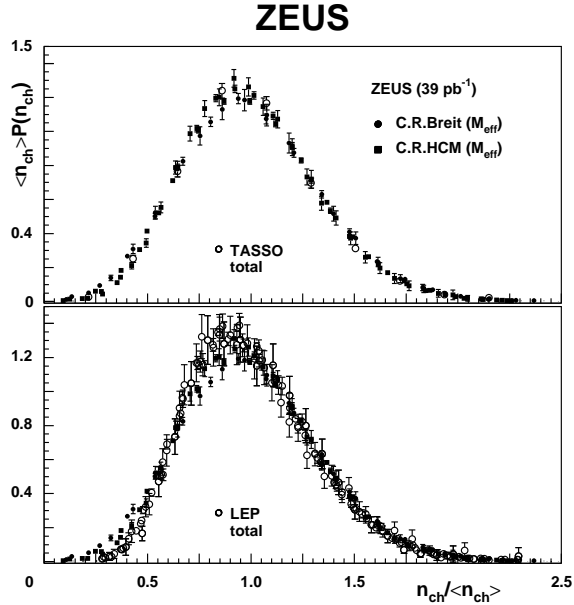}
    \end{center}
    \caption{KNO scaling of charged hadron multiplicity distributions measured in DIS events by ZEUS and compared to $\epem$ data taken by TASSO and LEP experiments. Taken from Ref.~\cite{Chekanov:2008hy}.}
    \label{fig:kno}
\end{figure}

Another successful check of universality was given by the recent high-precision ZEUS measurements of multiplicity distributions, $\dd{N}/\dd\nch$ at different scales~\cite{Chekanov:2008hy}, which compare very well with TASSO and LEP data. In particular, these data confirm that the product $\langle\nch\rangle\times\dd{N}/\dd\nch$ is a function of $\nch/\langle\nch\rangle$ independently of the jet energy, a property expected in QCD and known as Koba-Nielsen-Olesen (KNO) scaling~\cite{Koba:1972ng}, as can be seen in Figure~\ref{fig:kno}.

\subsubsection{Hadroproduction}

Because of the structure of both the projectile and the target, inclusive momentum hadron distributions in hadronic collisions may {\it a priori} reveal little on fragmentation processes, since for instance no scaled-momentum distributions such as Eq.~(\ref{eq:momdist}) can be constructed. However, it happens that the recent high-precision measurements of the BRAHMS \cite{Arsene:2007jd}, PHENIX~\cite{Adler:2003pb,Adare:2007dg}, and STAR~\cite{Adams:2006nd,Adams:2006uz,Abelev:2006cs} collaborations in $p$--$p$ collisions at RHIC, and CDF~\cite{Abe:1988yu,Acosta:2005pk} in $p$--$\bar{p}$ collisions at the Tevatron, actually allow for additional constraints to be set on FF parametrizations (see discussion in the next Section~\ref{sec:comparingsystems}).

The transverse momentum spectra of pions, kaons and (anti)protons at mid-rapidity\footnote{Measurements at large positive rapidity, $y\simeq3$, are also reported by BRAHMS~\cite{Arsene:2007jd} and STAR~\cite{Adams:2006uz}.} were measured in those experiments with a high accuracy up to roughly $\pt\simeq10$~GeV, and even up to $\pt\simeq20$~GeV for neutral pions by the PHENIX collaboration~\cite{Adare:2007dg}. Perhaps even more interesting are the measurements of strange particles such as $K_s^0$~\cite{Arsene:2007jd,Adams:2006nd} and $\Lambda$/$\bar{\Lambda}$~\cite{Abelev:2006cs}. Finally, the RHIC experiments BRAHMS and STAR were able to determine spectra differences between positive and negative-charged pions, $\Delta\pi^\pm=\dd\sigma(pp\to\pi^+\X)-\dd\sigma(pp\to\pi^-\X)$, a quantity obviously vanishing in $\epem$ annihilation and in $p$--$\bar{p}$ collisions at mid-rapidity. Because of the wealth of new measurements available from RHIC, those data sets are now systematically included in the recent global fit analyses DSS~\cite{deFlorian:2007aj} and AKK08~\cite{Albino:2008fy} (but not HKNS~\cite{Hirai:2007cx}) discussed in Section~\ref{sec:globalfits}. Before these data, the inclusive particle spectra in hadron collisions taken e.g. at the CERN S$p\bar{p}$S by the UA1 ~\cite{Albajar:1989an} and UA2~\cite{Banner:1983jq} collaborations were only used to test {\it a posteriori} the fragmentation functions extracted from $\epem$ data~\cite{Kniehl:2000fe,Kretzer:2000yf}, with the exception of the BFGW set~\cite{Bourhis:2000gs} which was actually the first analysis to consider hadroproduction measurements in order to get better constraints on the gluon fragmentation functions.

Single inclusive charged hadron spectra inside jets have also been measured very precisely in $p$--$\bar{p}$ collisions at the Tevatron by the CDF experiment~\cite{Acosta:2002gg} on a large variety of dijet masses, ranging from 80 to 600~GeV. Finally, let us note that jets of energies up to 50~GeV have been reconstructed lately by the STAR collaboration at RHIC~\cite{Abelev:2006uq}. This will eventually allow for single momentum spectra to be measured down to rather small values of $x$ and for different hadron species which was not done at the Tevatron. Preliminary measurements have recently been shown~\cite{Heinz:2008pc} which shapes seem to agree qualitatively well with the MLLA expectations discussed below. 

\subsubsection{Comparing systems}\label{sec:comparingsystems}

After having discussed the various experimental measurements related to fragmentation processes in $\epem$, $e(\nu)p$, $pp$, and $p\bar{p}$ collisions, it is worth to discuss the respective advantages and drawbacks of each collision system.

Electron-positron annihilation is the cleanest process and seems by far the most suited reaction to study fragmentation functions. Indeed, to leading order in the strong coupling $\alphas$, the momentum distribution $F^h(x, s)$,  Eq.~(\ref{eq:momdist}), measured in $\epem$ is simply related to the individual parton-to-hadron fragmentation functions, $D_i^h(x, Q^2)$, entering the fixed-order QCD calculation:~\cite{Ellis:1991qj}
\begin{equation*}
F^h(x, s) = \sum_{i=q,\bar{q}}\ g_i(s)\ D_i^h(x, s).
\end{equation*}
where $g_i(s)$ is the electroweak coupling. In analogy with the space-like case, the observable $F^h(x, s)$ is reminiscent of the structure function $F_2(x, Q^2)$ measured in DIS, and the individual FF $D_i^h(x, s)$ the analog of parton distribution functions, $f_i^h(x, Q^2)$. Beyond the leading order, factorization theorems~\cite{Collins:1981uw} ensure that $F^h$ can be expressed as
\begin{equation}\label{eq:epemnlo}
F^h(x, s) = \sum_{i=q,\bar{q},g}\ \int_x^1\ \frac{\dd{z}}{z}\ C_i\left(\frac{x}{z}, s, \mf^2\right)\ D_i^h\left(z, \mf^2\right),
\end{equation}
up to higher-twist terms which are power suppressed, e.g. $\cO{\lqcd^2/s}$. Both the parton-to-hadron FF, $D_i^h$, and the so-called coefficient functions, $C_i$, are factorization-scheme dependent (e.g. $\msbar$ or DIS) and depend on an arbitrary fragmentation scale, $\mf$, taken to be $\cO{\sqrts}$ in order to avoid large logarithmic corrections, $\ln(\mf^2/s)$. The coefficient functions have been computed in the $\msbar$ scheme a long time ago at NLO~\cite{Altarelli:1979kvBaier:1979sp} and more recently at NNLO~\cite{Rijken:1996ns}.

On top of its simplicity another obvious asset of $\epem$ scattering is the wealth of data available, as discussed in Section~\ref{sec:epem}. Moreover, the fragmentation of heavy quarks into either light-quark or heavy-quark hadrons is not as suppressed (at least as long as $m_Q\ll\sqrts$) as e.g. in DIS processes. These various reasons explain easily why almost only $\epem$ data have been used, until recently, to constrain FF parametrizations.

Despite those attractive features, however, $\epem$ collisions constrain pretty weakly the gluon FF which appear at NLO, as can be seen in Eq.~(\ref{eq:epemnlo}). Also, since hadrons produced in $\epem$ come from the fragmentation of both the quark {\it and} the anti-quark jet (at LO), those data only allow for the extraction of flavour-inclusive FF, e.g. $D_q^h+D_{\bar{q}}^h$ ($=D_q^{h/\bar{h}}$ from charge conjugation). This is clearly at variance with DIS and $p$--$p$ collisions in which quarks are produced {\it more abundantly} than anti-quarks in the hard process because of the valence-quark distribution in the (projectile and) target hadron --~at least as long as the momentum-fraction at which PDF are probed is not too small, say $x_{\rm Bj}\gtrsim0.1$. Therefore, the individual $D_q^h$ and $D_{\bar{q}}^h$ shall be determined independently from DIS and $p$--$p$ measurements\footnote{In principle, rapidity asymmetries in $p$--$\bar{p}$ collisions could also be used.}.

As already discussed in Section~\ref{sec:dis}, the advantage of DIS over hadroproduction experiments is the close connection offered with $\epem$ scattering, which allows one to observe the universality of fragmentation processes. Also, fragmentation functions can be probed on a wide $(x, Q^2)$-range unlike in $\epem$ for which the hard scale is set at LO by the centre-of-mass energy of the collision, $Q=\sqrt{s}/2$. Note however that at NLO, the hard scale also depends on the typical angle between two neighbouring jets $\theta$, $Q=\cO{E_{\rm jet}\sin(\theta/2)}$~\cite{Dokshitzer:1988me,Dokshitzer:1991wu} which allows for the variation of the hardness $Q$ in 3-jet events in $\epem$ collisions~\cite{Barate:1997ey}.

Coming to hadroproduction experiments, $p$--$p$ and $p$--$\bar{p}$ scattering are ideal in order to constrain the gluon fragmentation functions, especially when $x_{\rm Bj}\ll1$ hence when the hadron $\pt$ is small as compared to $\sqrts$. Note also that FF extracted in hadronic collisions are typically probed at {\it large} values of $x$ (typically $x\simeq0.7$--$0.8$ at RHIC~\cite{Kretzer:2004ie}) because the parton densities fall steeply when $x_{\rm Bj}=\cO{\pt/x}$ is large. Therefore the kinematic region probed in hadronic collisions is different, and hence complementary, to the one in $\epem$ collisions. Finally, measuring hadrons inside jets makes it possible to compare scaled-momentum spectra in hadroproduction experiments together with $\epem$ and DIS and check further the universality of fragmentation (see e.g. Figure~\ref{fig:coherence} below).

\subsection{Global fit analyses}\label{sec:globalfits}

The many attempts to extract FF from $\epem$ data have already been mentioned in the Introduction. We would like in this Section to focus only on the three most recent parametrizations, namely HKNS~\cite{Hirai:2007cx}, DSS~\cite{deFlorian:2007aj}, and AKK08~\cite{Albino:2008fy} --~which therefore should be used when computing inclusive hadron production in elementary collisions~-- emphasizing in particular the different assumptions made in each global fit analysis. For clarity, the assumptions and ingredients of each set are summarized in Table~\ref{tab:assumptions}.

\begin{table}[htdp]
\caption{Various characteristics of recent global fits analyses AKK08~\cite{Albino:2008fy}, DSS~\cite{deFlorian:2007aj} (and DSV~\cite{deFlorian:2007aj} for strange particles), and HKNS~\cite{Hirai:2007cx}. For each set we list the data samples used, the strange particles available, the error analyses, and further theoretical details.}
\begin{center}
\begin{tabular}{p{2.4cm}p{2.4cm}p{2.4cm}}
\hline 
\hline\\[-0.25cm]
\centerline{AKK08} & \centerline{DSS/DSV} & \centerline{HKNS} \\[-0.25cm]
\hline\\[-0.15cm]
\centerline{$\epem\bigm/pp\bigm/ {p\bar{p}}$} & \centerline{$\epem\bigm/ {e^-p}\bigm/ pp$} & \centerline{$\epem$} \\[-0.25cm]
\centerline{$K_s^0$, $\Lambda+\bar{\Lambda}$} & \centerline{$K_s^0$, $\Lambda+\bar{\Lambda}$} & \centerline{---}  \\[-0.25cm]
\centerline{---} & \centerline{Lagrange errors} & \centerline{{Hessian errors}} \\[-0.25cm]
\centerline{mass effects} & \centerline{---} & \centerline{---} \\[-0.25cm]
\centerline{large-$x$ resum.} & \centerline{---} & \centerline{---} \\[-0.25cm]
\hline
\hline
\end{tabular}
\end{center}
\label{default}
\label{tab:assumptions}
\end{table}

\subsubsection{Data samples}

As far as data samples are concerned, all three sets used almost the same $\epem$ measurements taken at LEP/SLC as well as at lower energy. Therefore, differences arise rather on the inclusion or not of DIS and hadroproduction data. While HKNS considered $\epem$ only, both AKK08 and DSS included the RHIC results (BRAHMS, PHENIX, STAR) to get a better handle on gluon FF; AKK08 also taking into account measurements of $K_s^0$ and $\Lambda/\bar{\Lambda}$ spectra performed at $\sqrts=630$~GeV by CDF~\cite{Acosta:2005pk}. DSS is the only analysis which included semi-inclusive DIS measurements by HERMES~\cite{Hillenbrand:2005pc}, despite the rather low $Q^2\sim1$--$5$~GeV$^2$ probed in that experiment.

\subsubsection{QCD evolution}

The DGLAP equations~\cite{Gribov:1972ri,Altarelli:1977zs,Dokshitzer:1977sg} governing the $Q^2$ evolution of parton-to-hadron fragmentation functions read 
\begin{equation}\label{eq:dglap}
\frac{\dd D_i(x, Q^2)}{\dd\ln Q^2} = \sum_j \int_x^1\ \frac{\dd{z}}{z}\ P_{ji}\left(\frac{x}{z},\alphas(Q^2)\right)\ D_j(z, Q^2)
\end{equation}
where $P_{ji}$ are the NLO time-like splitting functions determined in~\cite{Curci:1980uw,Furmanski:1980cm}. These evolution equations are either solved through a ``brute-force'' $z$-integration (HKNS) of~(\ref{eq:dglap}), or using the standard Mellin techniques (AKK08, DSS) writing Eq.~(\ref{eq:dglap}) as
\begin{equation*}
\frac{\dd {\tilde D}_i(N, Q^2)}{\dd\ln Q^2} = \sum_j\ {\tilde P_{ji}}\left(N,\alphas(Q^2)\right)\times {\tilde D}_j(N, Q^2)
\end{equation*}
in momentum space where the moment
\begin{equation*}
{\tilde D}_i(N, Q^2)=\int_0^1 \frac{\dd{z}}{z}\ z^N D_i(z, Q^2).
\end{equation*}
can then be Mellin-inverted to obtain the fragmentation function $D_i(x, Q^2)$,
\begin{equation*}
D_i(x, Q^2)=\frac{1}{2i\pi} \int_{{\cal C}} \dd{N}\ x^{-N} {\tilde D_i(N, Q^2)},
\end{equation*}
where ${\cal C}$ is a contour  which lie at the right of all singularities. The reader may refer to~\cite{Kumano:2004dw} for a discussion on the various approaches. Mellin techniques also proved useful in computing the hadroproduction cross sections at NLO in a very efficient way. Indeed, the Mellin-transform of the hadronic coefficient functions $C_{i,j}^k(z)$,~i.e. complicated functions involving the product of two parton densities and the partonic QCD cross sections, need only to be computed once~\cite{deFlorian:2007aj}. Therefore the hadron-hadron production cross section, symbolically written as
\begin{equation*}
E_h\frac{\dd^3\sigma}{\dd^3p_h}=\sum_{i,j,k}\ \int_x^1\frac{\dd{z}}{z}\ C_{i,j}^k(x/z)\times D_k^h(z)=\frac{1}{2i\pi}\sum_k\int_{\cal C}\dd{N} \ \tilde{D}_k^h(N)\times \underbrace{\sum_{i, j} \tilde{C}_{i,j}^k(N)}_{\rm tabulated},
\end{equation*}
can easily be determined, exploiting the fact that the moments of the FF drop rapidly with $N$,  $\tilde{D}_k^h(N)\propto N^{-\beta}$~\cite{deFlorian:2007aj}, where $\beta$ is the large-$x$ power law of the FF, $D_k^h(z)\propto (1-z)^{\beta}$.

\subsubsection{Fit assumptions}\label{sec:fitassumptions}

All groups assume SU(2) isospin symmetry for the sea (also called {\it unfavoured}) fragmentation functions $D_i^h$, that is processes for which the hadron $h$ does not contain the flavour $i$ as a valence parton. Taking the $\pi^+=|u\bar{d}\rangle$ as an example, they assume that
\begin{equation*}
D_{\bar{u}}^{\pi^+}=D_d^{\pi^+}.
\end{equation*}
HKNS assume moreover that $D_{s}^{\pi^+}=D_{\bar{u}}^{\pi^+}$, yet the SU(3) isospin symmetry could be pretty badly broken because of the strange quark mass; DSS report for instance $D_{s}^{\pi^+}=0.83 D_{\bar{u}}^{\pi^+}$ from their fit. In the valence (favoured) sector, HKNS and AKK08 conjecture that SU(2) symmetry also holds, $D_{u}^{\pi^+}=D_{\bar{d}}^{\pi^+}$, while DSS allow for the different normalization of the valence plus sea quark FF, $D_{d+\bar{d}}={\cal N}\times D_{u+\bar{u}}$, ${\cal N}$ being a free parameter which is not necessarily equal to one (as assumed by AKK08 and HKNS). From their global fit analysis, and in particular thanks to the SIDIS HERMES preliminary measurements, DSS obtained ${\cal N}=1.10$, yet the value of ${\cal N}$ depends somehow on the individual fitted data samples. They also check the assumption made by Kretzer~\cite{Kretzer:2000yf} that unfavoured fragmentation functions are suppressed at large $x$ by one extra-power in $(1-x)$ with respect to favoured ones, e.g. $D_{\bar{u}}^{\pi^+}(x)=(1-x)D_u^{\pi^+}(x)$. This working hypothesis turns out to be in rather good agreement with $\pi^+$ data from HERMES~\cite{Hillenbrand:2005pc}, although it somehow failed to describe $\pi^-$ production, $D_u^{\pi^-}(x)\ne(1-x)D_{\bar{u}}^{\pi^-}(x)$~\cite{deFlorian:2007aj}.

The analytic parametrization of the FF at an initial scale $Q_0$ (1~GeV$^2$ for DSS and HKNS, 2~GeV$^2$ for AKK08) is
\begin{equation}\label{eq:param}
D_i^{h^\pm}(x, Q_0^2) = N_i^{h^\pm} x^{a_i^{h^\pm}} (1-x)^{b_i^{h^\pm}} \left[ 1-c_i^{h^\pm} (1-x)^{d_i^{h^\pm}}\right]
\end{equation}
in AKK08 and DSS, and taking the simpler form $c_i^{h^\pm}\equiv0$ in HKNS. The normalization parameters $N_i^{h^\pm}$ are constrained in such a way to satisfy the following momentum sum rule,
\begin{equation}\label{eq:sumrule}
\sum_h \int_0^1\ \dd{z}\ z D_i^h(z, Q^2) = \sum_h \tilde{D}_i^h(2, Q^2) = 1,
\end{equation}
for all parton species $i$ and all $Q^2$. Rather than giving the FF into charge-sign identified hadrons, e.g. $D_i^{\pi^+}$ and $D_i^{\pi^-}$, AKK08 chooses another ``basis'', the inclusive-charged $D_i^{\pi^++\pi^-}$ as well as the ``valence'' quark FF, $\Delta_i^{\pi^\pm}=D_i^{\pi^+}-D_i^{\pi^-}$, which is needed to compute charge-sign asymmetries e.g. at RHIC or in semi-inclusive DIS.

\begin{figure}[htb]
    \begin{center}
      \includegraphics[width=9.cm]{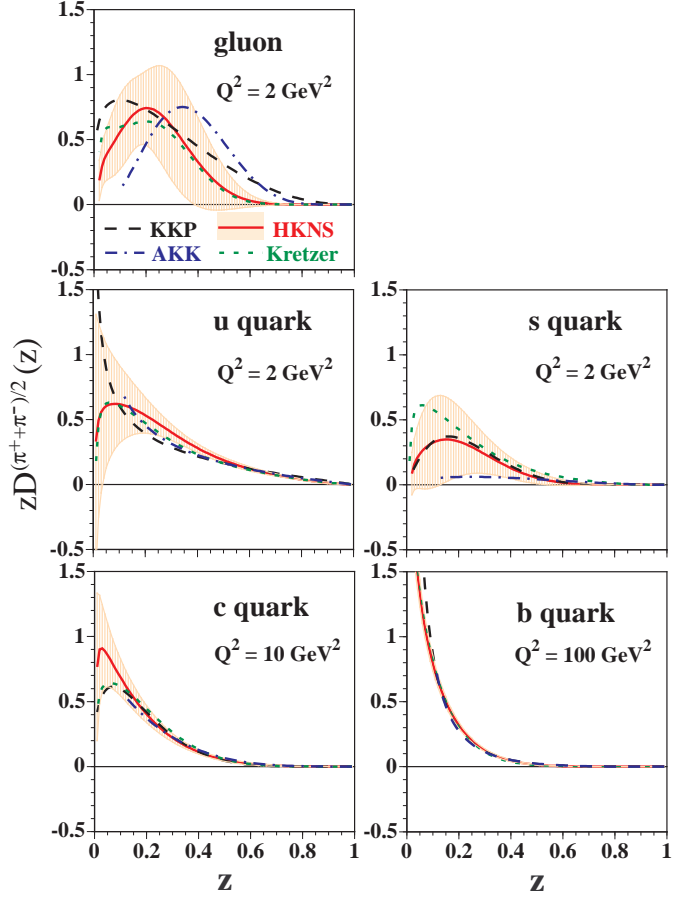}
    \end{center}
    \caption{HKNS parametrization for charged pions, $D_i^{\pi^\pm}/2$, at NLO for all flavours, compared to the AKK05, KKP, and Kr sets previously available. The bands indicate the theoretical uncertainty of the HKNS fragmentation functions. Taken from Ref.~\cite{Hirai:2007cx}.}
    \label{fig:hkns}
\end{figure}

\subsubsection{Error analyses}

Besides the use of new data sets available from RHIC for instance, the most important improvement of the recent FF sets concerns the treatment of errors in the global fit analyses. Fragmentation studies have in particular greatly benefited from the progress made in the extraction of parton distribution functions over the last few years~\cite{Pumplin:2001ct,Stump:2001gu,Botje:1999dj}.

Using the Hessian method~\cite{Pumplin:2001ct} which assumes a parabolic dependence of the $\chi^2$ function for parameter values away from its minimum, HKNS compute the spread of the FF propagating the (correlated) errors on the individual parameters entering Eq.~(\ref{eq:param}). The advantage of this error analysis is to make rapidly clear which fragmentation functions are the most uncertain and in which kinematical domain\footnote{Note also that a \texttt{FORTRAN} code released by HKNS allows for computing the theoretical uncertainty coming from their FF set on any observable.}. For the illustration, the HKNS individual NLO fragmentation functions into charged pion data are displayed in Figure~\ref{fig:hkns} and compared to the ``older'' sets AKK05, KKP, and Kr; the envelopes of the FF indicate the theoretical error from the Hessian analysis.

DSS used instead the Lagrange multiplier technique~\cite{Stump:2001gu} which helps to understand the range of variation of physical observables allowed from the global fit, taking the truncated second moment as an example in~\cite{deFlorian:2007aj}. Also, interestingly, the influence of a specific data sample on a given observable can be easily determined within this approach.

\subsubsection{Hadron mass corrections and large-$x$ resummation}

Unlike DSS and HKNS, AKK08 have not performed an error analysis, although work towards this goal is in progress. What is performed in AKK08 and not carried out elsewhere, however, is the treatment of hadron mass effects, $\mh\ne0$, in the extraction of fragmentation functions. They are taken into account by making explicit the kinematical differences between the light-cone scaling variable $x$ entering the fragmentation functions and the energy ($x_E$) or momentum ($x_p$) fraction used experimentally e.g. in $\epem$ collisions (corrections can be taken care of in a similar manner in hadronic collisions):
\begin{equation*}
x_{E, p} = x\ \left(1\pm\frac{\mh^2}{x^2 s}\right).
\end{equation*}
The AKK08 fits have been performed for all hadron species letting $\mh$ as a free parameter. Remarkably, the fitted mass turned out to be amazingly close to the true masses (1\% difference is quoted) for baryons ($p$/$\bar{p}$, $\Lambda$/$\bar{\Lambda}$), yet 10\% larger for pions, an observation attributed in~\cite{Albino:2008fy} to the decay of the $\rho$-meson, whose mass is much larger than that of the pion; note however that the fitted kaon mass proves 30\% smaller than its actual value. On the phenomenological side, hadron mass effects are visible at low $x$ and/or small $\sqrts$.

The other novelty of the AKK08 analysis is the large-$N$ (large $x$) resummation of leading logarithms, $\alphas^n\ln^{n+1}N$, and next-to-leading logarithms, $\alphas^n\ln^nN$, of the Mellin transform of the NLO $\epem$ coefficient functions, $C_i$, appearing in Eq.~(\ref{eq:epemnlo})~\cite{Cacciari:2001cw}. Large-$x$ logarithms appearing in the DGLAP evolution are also resummed to LL and NLL accuracy, using~\cite{Albino:2007ns}. Even though the inclusion of large-$x$ resummation improves the fit for $K^\pm$, $p/\bar{p}$, and especially $\Lambda/\bar{\Lambda}$ production ($\chi^2/{\rm ndf}=1.45$ instead of $\chi^2/{\rm ndf}=1.73$ without resummation), differences with the purely fixed-order analysis are however pretty small, and only visible either at low energy or above $x\gtrsim0.7$. Given that fragmentation functions are extracted within a resummed analysis, one may also wonder whether AKK08 should in principle be used in a fixed-order calculation.

\subsubsection{Comparing FF sets}

A detailed comparison of the various sets would not be appropriate here; the reader is rather referred to the original publications of each analysis~\cite{Albino:2008fy,deFlorian:2007aj,Hirai:2007cx} for a complete discussion. The general features are thus only briefly sketched in this Section. First of all, given the amount of $\epem$ data available, the fragmentation of light quarks into $\pi^\pm$ is rather similar in each set. However, the gluon fragmentation functions, $D_g^{\pi^\pm}$, reported by AKK05/AKK08 and especially that of DSS (both using hadronic data) are much harder than the one given by HKNS which is only extracted from $\epem$ measurements. This can be seen in the upper left panel of Figure~\ref{fig:hkns} where AKK05 and HKNS are compared. Therefore, potentially large differences in the QCD predictions of hadron $\pt$-spectra e.g. at the LHC could be expected, AKK08 and DSS being {\it a priori} more trustful than HKNS because of the additional constraints brought by RHIC and Tevatron data. Nevertheless, the slight disagreement between AKK08 and DSS found at high $x$ is a hint that systematic theoretical uncertainties on the $g\to\pi^\pm$ process may still be large. Indeed, even though the hadronic measurements help to constrain gluon fragmentation processes, the data in $p$--$p$ and $p$--$\bar{p}$ reactions are not as precise as the $\epem$ data samples, and uncertainties remain substantial at high values of $x$. Even though the agreement between data and theory may vary from one set or one hadron species to another, all NLO predictions of $\pi^\pm$ and $K^\pm$ spectra reproduce fairly well the available data. Remarkably, those sets prove also rather successful in predicting baryon production such as $p/\bar{p}$ (despite the agreement is not as good as for $\pi^\pm$), yet the FF into $p/\bar{p}$ are rather dissimilar in each set. In the case of $\Lambda/\bar{\Lambda}$ production, however, a large discrepancy between theory and STAR data in $p$--$p$ collisions~\cite{Abelev:2006cs} is apparent either when using AKK08~\cite{Albino:2008fy} or DSV~\cite{deFlorian:2007aj} sets, which both undershoot the data by roughly a factor of 5. Finally, the largest differences between the parametrizations are seen for the charge-asymmetry performed at RHIC which should eventually be able, when the data become more precise, to discriminate the various assumptions regarding favoured {\it vs.} unfavoured fragmentation functions.

While it is a common procedure to vary the renormalization/factorization in order to give an estimate of unknown higher order corrections, that is the perturbative error, it is a clear that the predictions of hadron spectra should ideally also use all the three recent sets AKK08, DSS, and HKNS, in order to quantify the ``non-perturbative uncertainty'' of these calculations. Having a tool similar to the recent LHAPDF interface for parton densities~\cite{Giele:2002hx} would of course help significantly in this respect.

\subsubsection{Towards NNLO analyses?}

All the global fit analyses performed recently showed that the agreement between the FF sets and the current data was (much) higher when going from LO to NLO accuracy. Therefore it is legitimate to wonder whether even higher-order corrections would improve further the situation or not.

The calculation of the space-like splitting functions at three loops, $\cO{\alphas^3}$, has been carried out a few years ago by Moch, Vogt and Vermaseren~\cite{Moch:2004pa}, then allowing for the first analyses of parton densities at NNLO accuracy~\cite{Martin:2002dr}. Unfortunately, no complete calculation is yet available at this order for time-like splitting functions. At leading order, $\msbar$ space-like and time-like evolution kernels are identical, an observation known as Gribov-Lipatov reciprocity~\cite{Gribov:1972rt,Gribov:1972ri}. Beyond that order time-like splitting functions can be determined from an analytic continuation in $x$~\cite{Furmanski:1980cm,Stratmann:1996hn} or through the conjecture made by Dokshitzer, Marchesini and Salam~\cite{Dokshitzer:2005bf} who suggest a simple correspondence between space-like and time-like evolution to all orders from a redefinition of the standard DGLAP equations\footnote{Basically, the scale at which the PDF or FF must be evaluated in the r.h.s. of (\ref{eq:dglap}) is $Q^2 z$ and $Q^2/z$, respectively, with new ``universal'' splitting functions ${\cal P}(z, \alphas(Q^2/z))$.}.

Based on those two proposals, Mitov, Moch and Vogt recently achieved the computation of the NNLO non-singlet splitting functions~\cite{Mitov:2006ic}. There is therefore a real hope that analyses for fragmentation functions at NNLO can be performed in the near future.

\subsection{Fragmentation at small $x$}\label{sec:mlla}

The DGLAP evolution~(\ref{eq:dglap}) deals with large collinear logarithms which are resummed within the Leading Logarithmic  Approximation (LLA). As a consequence of {\it strong} transverse momentum ($k_\perp$) ordering in the evolution, all powers of hard collinear logarithms, $\alpha_s\ln\kt^2$ with $\kt/\omega\ll1$, are resummed. This scheme describes the space-like and time-like evolution of the parton densities and fragmentation functions, respectively, which both satisfy DGLAP equations. In the literature it is also called the ``fixed order approach''. However, when radiated gluons in the process reach the infrared region (that is, $x\ll1$), the resummation of large soft logarithms becomes also necessary, which leads to the Double Logarithmic Approximation (DLA)~\cite{Bassetto:1982ma,Dokshitzer:1991wu}. 

In DLA, as a consequence of {\it strong} angular ordering (AO) in cascading processes, all powers of soft and collinear logarithms, $\alphas\ln(1/x)\ln\kt^2$ are resummed. This
scheme describes the time-like evolution of fragmentation functions and takes properly into account the crucial QCD coherence effects: soft and collinear gluons interfere destructively once $k_\perp$ becomes smaller than a certain value. In addition, it was demonstrated that the dominant contribution to inclusive particle production is obtained when the successive emissions of soft and collinear gluons are strongly ordered in their energies 
($\omega_i\gg\omega_{i+1}$) and emission angles ($\vartheta_i\gg\vartheta_{i+1}$).
This hierarchy in energies and angles in DLA gives rise to a simple probabilistic interpretation in terms of classical parton shower cascades, which can easily be implemented in Monte Carlo approaches. Indeed, within this picture, branching processes in a high energy jet can be displayed in terms of Feynman diagrams at tree level, which tremendously simplifies the problem by getting rid of all possible interferences. However, in this approximation the energy balance at each vertex is ignored (basically assuming $1-x\simeq1$), which leads to an overestimate of particle production in DLA. 

Adding the double logarithms (DLA) to the single collinear (but hard) leading logarithms (LLA) leads to the Modified Leading Logarithmic Approximation (MLLA) proposed in Refs.~\cite{Dokshitzer:1991wu,Dokshitzer:1982fh,Azimov:1985by}. As compared to DLA, MLLA partially restores the energy balance  for each gluon emission --~although it remains approximate~-- and takes explicitly into account the running of the coupling constant in the evolution, from the hardest scale $Q$ to threshold $Q_0$. Remarkably, the probabilistic interpretation encountered in DLA is preserved in MLLA with a simple prescription on the angular ordering, which now becomes {\it exact}: $\vartheta_i\geq\vartheta_{i+1}$, after azimuthally averaging over the phase space of the emitted gluon. Consequently, all powers of $\alphas\ln(1/x)\ln\vartheta\ +\ \alphas\ln\vartheta$ are resummed in the MLLA, where
$\alphas\ln(1/x)\ln\vartheta=\cO{1}$ and $\alphas\ln\vartheta=\cO{\sqrt{\alphas}}$. The collinear terms are known to partially restore the energy balance that is overestimated in the DLA.

\begin{figure}[htb]
    \begin{center}
      \includegraphics[width=7.5cm]{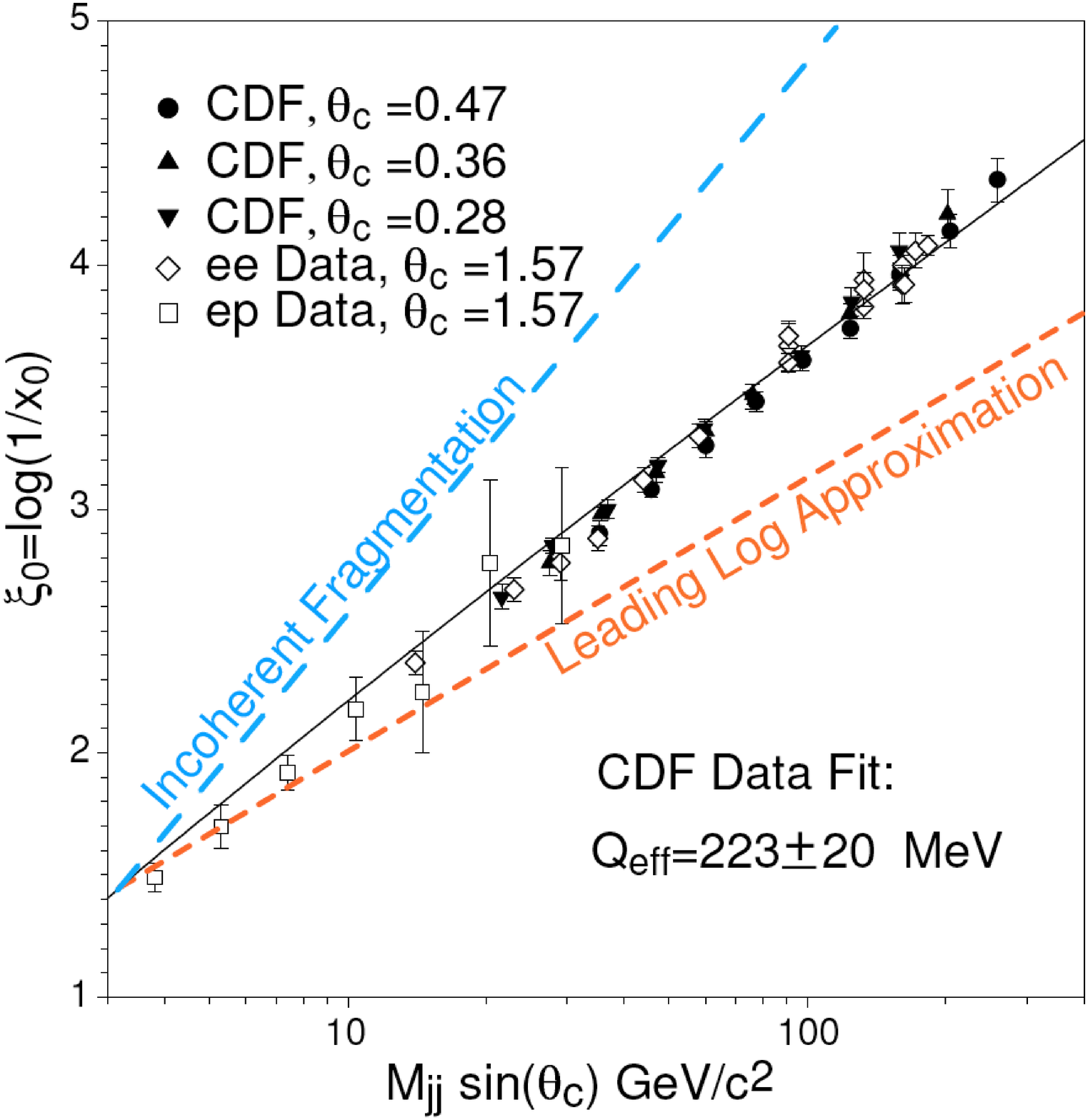}
    \end{center}
    \caption{Location of the peak of the single-inclusive distribution $xD(x,Q^2)$ as a function of the energy scale $Q$, measured in $\epem$, DIS, and $p$--$\bar{p}$ collisions. The prediction in the Leading-Logarithmic Approximation (short-dashed line), MLLA (solid line), and neglecting coherence effects (long-dashed line) are shown for comparison. Figure taken from Ref.~\cite{Acosta:2002gg}.}
    \label{fig:coherence}
\end{figure}

The small-$x$ evolution equation can be obtained from the DGLAP evolution equation with an appropriate change of the evolution variable, from the virtual mass-squared $Q^2$ to $z^2\ Q^2$, i.e.
\begin{equation}\label{eq:dglapsmallx}
\frac{\dd D_i(x, Q^2)}{\dd\ln Q^2} = \sum_j \int_x^1\ \frac{\dd{z}}{z}\ P_{ji}\left(\frac{x}{z},\alphas(Q^2)\right)\ D_j(z, z^2 Q^2),
\end{equation}
which reduces to Eq.~(\ref{eq:dglap}) at $x=\cO{1}$. Because of this change of scale in (\ref{eq:dglapsmallx}), the anomalous dimensions $\gamma_{ji}$ are no longer divergent at $j=1$ (which corresponds to $x=0$). In the $gg$ case, it is given by~\cite{Ellis:1991qj},
\begin{equation}\label{eq:anomdimsmallx}
\gamma_{gg}(j,\alphas)=\alphabar\ \frac{2}{j-1 +2 \gamma_{gg}(j, \alphas)},
\end{equation}
with $\alphabar\equiv\alphas N_c/ 2\pi$. Performing the Taylor expansion of (\ref{eq:anomdimsmallx}) in $(j-1)^2/\alphabar\ll1$,~\cite{Ellis:1991qj}
\begin{equation}\label{eq:anomdimsmallxtaylor}
\gamma_{gg}(j,\alphas)=\sqrt{\alphabar}-\frac14(j-1)+ \frac{1}{32 \sqrt{\alphabar}}(j-1)^2 + {\cal O}\left(\frac{(j-1)^4}{{\alphabar}^{3/2}}\right),
\end{equation}
allows one to extract the leading small-$x$ contribution to multiplicities or particle spectra. Note that, interestingly, $\gamma_{gg}\propto\sqrt{\alphas}$ at small-$x$. On the contrary $\gamma_{gg}$ can be written as a power series in $\alphas$ when Taylor expanding (\ref{eq:anomdimsmallx}) at small coupling constant (and/or large $j-1$),
\begin{equation}
\gamma_{gg}(j,\alphas)=\alphabar\ \frac{1}{j-1}-\alphabar^2\frac{2}{(j-1)^3} + {\cal O}\left(\frac{\alphabar^3}{(j-1)^5}\right).
\end{equation}
From the anomalous dimension (\ref{eq:anomdimsmallxtaylor}) the moments of the fragmentation function, $\tilde{D}(j, t)\sim t^\gamma$, can be obtained~\cite{Ellis:1991qj},
\begin{equation}\label{eq:dtildedla}
\tilde{D}(j, Q^2)\sim\exp\Bigg[\frac1b\sqrt{\frac{2 N_c}{\pi\alphas}}-\frac{1}{4 b \alphas}(j-1)+ \frac{1}{48 b}\sqrt{\frac{2 \pi}{\alphas^3 N_c}}(j-1)^2+\cO{\frac{(j-1)^4}{\alphabar^{5/2}}}\Bigg],
\end{equation}
where $\alphas$ is evaluated at the scale $Q^2$ and $b=(11 N_c/3- 2n_f/3)/2\pi$. Performing the inverse Mellin transform of Eq.~(\ref{eq:dtildedla}) leads to the DLA fragmentation function,
\begin{equation}\label{eq:spectrumdla}
x D(x, Q^2) \sim \exp\left[-\frac{1}{2\sigma^2} (\xi-\xi_p)^2\right],
\end{equation}
which is Gaussian in $\xi=\ln(1/x)$. The distribution --~known as ``hump-backed plateau''~-- has a maximum because of gluon QCD coherence at small $x$ (large $\xi$), see Figure~\ref{fig:distortinghbp} below. The location of the peak and the width of the distribution are given by
\begin{eqnarray}
\xi_p&=&\frac{1}{4 b \alphas}=\frac12 \ln\left(\frac{Q}{\lqcd}\right),\\
\sigma^2&=&\frac{1}{24 b} \sqrt{\frac{2\pi}{\alphas^3 N_c}}=\frac{\sqrt{6}}{12} \ln^{3/2}\left(\frac{Q}{\lqcd}\right).
\end{eqnarray}
The DLA inclusive spectrum (\ref{eq:spectrumdla}) is correct at asymptotic energies. When including single logarithm corrections, the MLLA spectrum is softened as compared to DLA and its peak is shifted towards larger values of $\xi$ by an amount $\propto\sqrt{\alphas}$,~\cite{Dokshitzer:1991wu}
\begin{equation*}
\xi_p^{^{\rm MLLA}}= \ln\left(\frac{Q}{\lqcd}\right)\ \left[ \frac12 + a \sqrt{\frac{\alphas}{32 N_c \pi}} + \cO{\alphas} \right],
\end{equation*}
with $a\equiv11 N_c/3+2n_f/3N_c^2$. The MLLA spectrum is expressed in~\cite{Dokshitzer:1991wu} in terms of the inverse Mellin transform of a confluent hypergeometric function. Recently, an integral representation has been obtained by P\'erez~Ramos beyond the limiting spectrum approximation, that is when the infrared scale $Q_0\ne\lqcd$ ($\lambda\equiv\ln(Q_0/\lqcd)\ne0$),~\cite{Ramos:2006dx}
\begin{equation*}
G\left(\xi,y\right)=\left(\xi\!+\!y\!+\!\lambda\right)\!\!\iint
\frac{\dd\omega\, \dd\nu}{\left(2\pi i\right)^2}e^{\omega\xi+\nu y}
\ \int_{0}^{\infty}\frac{\dd s}{\nu+s}
\!\!\times\left(\!\frac{\omega
\left(\nu+s\right)}{\left(\omega+s\right)\nu}\!\right)^{1/\beta_0
\left(\omega-\nu\right)}\!\!\left(\!\frac{\nu}{\nu+s}\!\right)^
{a_1/\beta_0}\,e^{-\lambda s},
\end{equation*}
(with $y\equiv\ln(Q/\lqcd)-\xi$) which was then estimated analytically in Ref.~\cite{Ramos:2006mk} using the steepest descent method.

\begin{figure}[htb]
    \begin{center}
      \includegraphics[width=8.5cm]{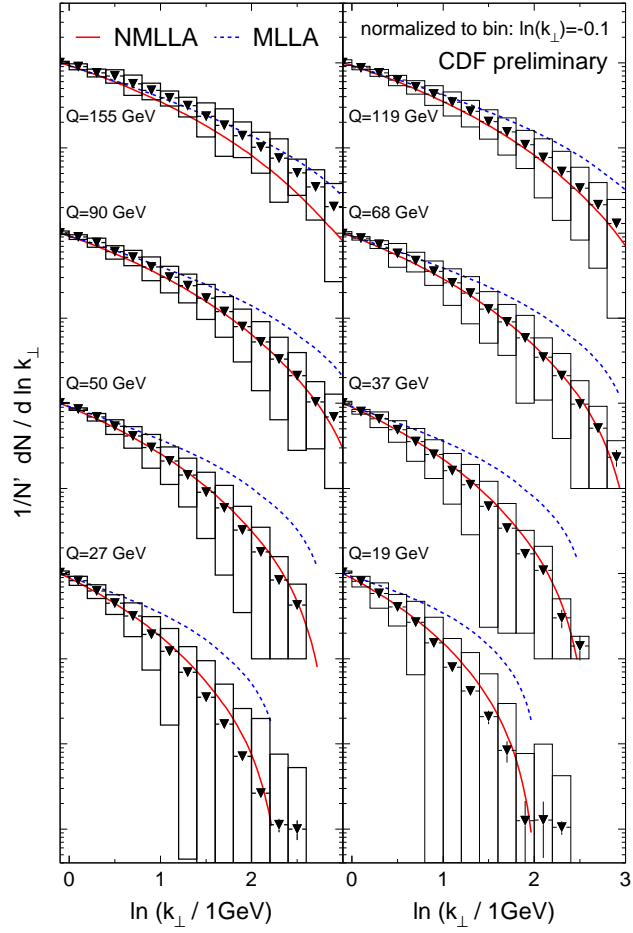}
    \end{center}
    \caption{CDF preliminary measurements~\cite{Jindariani:2006ye} of $\kt$-distributions for hadrons inside a jet compared to MLLA (dashed line) and NMLLA (solid line) calculations. Figure taken from Ref.~\cite{Arleo:2007wn}.}
    \label{fig:ktdist}
\end{figure}

The shape of the MLLA inclusive spectrum describes successfully the momentum distributions of hadrons measured in $\epem$, DIS and $p$--$\bar{p}$ collisions at the Tevatron (see e.g.~\cite{Khoze:1996dn} for a comprehensive review on MLLA phenomenology). In particular, the decrease of parton spectra at small $x$ (large $\xi$) because of QCD coherence is observed experimentally. Another check of the MLLA framework is the scale dependence of the location of the hump --~i.e. independent of the spectrum normalization~-- which compares well with the experimental results (Figure~\ref{fig:coherence}) and brings further evidence for coherence effects in QCD at small $x$.

Remarkably, apart from the value of $\lqcd$, the only parameter entering the MLLA spectrum is an overall normalization factor ${\cal K}$ of the order of 1. The similarity of the shape of {\it partonic} spectra computed in MLLA on the one hand and that of {\it hadrons}  measured experimentally on the other hand supports the idea of Local Parton Hadron Duality (LPHD) introduced in~\cite{Azimov:1984np,Azimov:1985by}: hadronization happens locally and thus does not distort dramatically the spectra of quarks and gluons when they turn into hadrons. 

More differential observables have been considered lately, such as the transverse momentum spectra of hadrons inside a jet, $\dd{N}/\dd\kt$, which has been computed for the first time by P\'erez~Ramos and Machet at MLLA and in the limiting spectrum approximation ($Q_0=\lqcd$)~\cite{PerezRamos:2005nh}. More recently, the calculation has been extended beyond the limiting spectrum and including $\cO{\alphas}$ corrections which are strictly speaking next-to-MLLA (NMLLA)~\cite{Arleo:2007wn}. On the experimental side, the CDF collaboration reported on preliminary measurements of $\kt$-distributions of inclusive-charged hadrons from dijet production in $p$--$\bar{p}$ collisions at $\sqrts=1.96$~TeV~\cite{Jindariani:2006ye}, on a large variety of scales, $Q=M_{jj}\sin(\theta_c/2)=19$--$155$~GeV, where $M_{jj}$ is the dijet invariant mass and $\theta_c$ the jet cone radius. As can be seen in Figure~\ref{fig:ktdist}, the NMLLA predictions reproduce very well the experimental measurements (both data and theory are normalized at $\ln(\kt/1\ {\rm GeV})=-0.1$) for all energy scales. Another interesting fact is that MLLA spectra prove harder than CDF data, which is an indication that recoil effects neglected in MLLA are necessary to account for the experimental results.

Recently, the small-$x$ evolution equations in the coherent branching formalism~\cite{Bassetto:1984ik} (of which Eq.~(\ref{eq:dglapsmallx}) is an approximation) has been solved numerically~\cite{Sapeta:2008km}, allowing for the calculation of single spectra with exact energy conservation in the evolution. Although no significant differences with MLLA are observed at large jet energies ($\ln(Q/Q_0)\gtrsim7$), the shape of the hump-backed plateau proves somehow modified at lower energy~\cite{Sapeta:2008km}.

Finally, another interesting development has been suggested by Albino, Kniehl, Kramer and Ochs~\cite{Albino:2005gg} in order to unify the standard large-$x$ DGLAP evolution equation, Eq.~(\ref{eq:dglap}), together with the DLA evolution at small values of $x$, Eq.~(\ref{eq:dglapsmallx}). The DGLAP equation Eq.~(\ref{eq:dglap}) is solved using new splitting functions which resum the double logarithms to all orders (DLA part) supplemented by the regular part $\overline{P}$ of the fixed-order splitting function in order to avoid any double counting:~\cite{Albino:2005gg}
\begin{equation*}
\alphas\ P^0(z) \to P^{\rm DL}(z, \alphas) + \alphas\ \overline{P}^0(z).
\end{equation*}
Remarkably, solving DGLAP equation with these modified splitting functions allows for an excellent description of single inclusive $\epem$ data at all values of $x$, from  very small $x$ ($\xi\simeq5$) and up to very high values, $x\simeq 0.9$. Moreover, the extrapolation from large to small $x$ does not require the help of LPHD and the inclusion of an arbitrary normalization factor.

\subsection{Fragmentation into heavy-quark hadrons}\label{sec:heavyquark}

Heavy quark production can be computed perturbatively because the large mass, $m_Q \gg \lqcd$, acts as a collinear cut-off for the gluon emission by a heavy quark. Therefore, the fragmentation function,  $D_i^Q(x, Q_0, m_Q)$, of a parton with flavour $i$ into a heavy quark and at an initial scale $Q_0$ can be computed order by order in perturbation theory. The calculation has been performed at $\cO{\alphas}$ in~\cite{Mele:1990cw}:
\begin{eqnarray}
D_Q^Q(x, Q_0^2, m_Q) &=& \delta(1-x) + \frac{\alphas C_F}{2\pi}\label{eq:dQQ} \left[\frac{1+x^2}{1-x} \left(\log\frac{Q_0^2}{(1-x)^2 m_Q^2}-1\right)\right]_+\\
D_g^Q(x, Q_0^2, m_Q)&=& \frac{\alphas T_F}{2\pi}\left[x^2+(1-x)^2\right]\log\frac{Q_0^2}{m_Q^2}\label{eq:dgQ}\\
D_{q, \bar{q}}^Q(x, Q_0^2, m_Q)&=& \cO{\alphas^2}\label{eq:dqlightQ}
\end{eqnarray}
with $C_F=4/3$, $T_F=1/2$. The $D_Q^Q$ fragmentation function receives a singular contribution corresponding to no gluon emission, while $D_g^Q$ only appears at order $\alphas$ from the $g\to{Q}\bar{Q}$ splitting process and $D_{q\bar{q}}^Q=\cO{\alphas^2}$ from the double cascade $q(\bar{q})\to q(\bar{q})g$ followed by $g\to{Q}\bar{Q}$. The initial scale is $Q_0=\cO{m_Q}$ in order to avoid large logarithms appearing in Eqs.~(\ref{eq:dQQ}) and (\ref{eq:dgQ}). Once these initial conditions are known, the FF can be evolved through LO or NLO DGLAP evolution, Eq.~(\ref{eq:dglap}), from $Q_0$ to the hard scale $Q$ of the reaction. This procedure ensures that the leading $\alphas^n(Q)\ln^nQ/Q_0$ and next-to-leading $\alphas^{n}(\mu)\ln^{n-1}Q/Q_0$ logarithms of collinear origin are resumed to all orders.

As can be seen from Eq.~(\ref{eq:dQQ}), the initial condition for the heavy-quark fragmentation function contains large terms, $\log(1-x)/(1-x)$ which are divergent in the $x\to1$ limit. These terms arise at any order in the perturbative expansion, because of the non-cancellation of real and virtual terms in gluon emission, and need to be resumed in order to obtain an accurate description of fragmentation in this kinematic limit. This task has been achieved first at leading logarithmic~\cite{Mele:1990cw} and then at next-to-leading order accuracy in~\cite{Dokshitzer:1995ev,Cacciari:2001cw}, through the inclusion of Sudakov form-factors in the initial condition~(\ref{eq:dQQ}).

As in the case of light hadron production, the heavy-quark scaled-momentum spectrum is given by
\begin{equation}\label{eq:FQ}
F^Q(x, s, m_Q) = \sum_i\ \int_x^1\ \frac{\dd{z}}{z}\ C_i\left(\frac{x}{z}, s, \mu^2\right)\ D_i^Q\left(z, \mu^2, m_Q\right).
\end{equation}
Even though the sum runs over all flavours $i$, it is worth to mention that in $\epem$ collisions heavy quark production from gluon splitting only contributes at a few percent level as indicated for charm production by the ALEPH data~\cite{Heister:2003bu} (see also~\cite{Mueller:1985zp} for theoretical studies), even though it is expected to be somewhat larger in hadronic collisions where gluons are produced at the leading order. Because of the large mass, the heavy quark carries most of the jet momentum and therefore the single inclusive spectra of heavy hadrons are peaked towards large values of $x$~\cite{Bjorken:1977md}, unlike the spectra of light hadrons that we have seen in Section~\ref{sec:datalight} are peaked at small $x$.

When computing heavy-flavoured hadrons, $H$, however, non-perturbative effects come into play because of the $Q\to H$ hadronization mechanism. Therefore the heavy-quark spectrum (\ref{eq:FQ}) has to be convoluted with a non-perturbative fragmentation function, $D_{\rm np}$, in order to give the heavy-hadron spectrum:
\begin{equation}\label{eq:FH}
F^H(x, s) = \sum_i\ \int_x^1\ \frac{\dd{z}}{z}\ F^Q\left(\frac{x}{z}, s, m_Q\right)\ D_{\rm np}\left(z\right),
\end{equation}
which in Mellin space simply reads
\begin{equation*}
{\tilde F}^H(N, s) = \sum_i\ {\tilde C}_i(N, s, \mu^2) {\tilde D}_i^Q(N, \mu^2, m_Q) {\tilde D}_{\rm np}\left(N\right).
\end{equation*}
Several analytic parametrizations of non-perturbative fragmentation functions have been given (summarized in Table~\ref{tab:npff}) which parameters can be constrained from a comparison with existing data.

{\footnotesize
\begin{table}[htb]
\begin{center}
\caption{
\label{tab:npff} 
Non-perturbative heavy-quark fragmentation functions, $D_{\rm np}(x)$. The $f_i$ appearing in the BCFY parametrization are polynomials of degree $i$. 
}
\begin{tabular}{lc}
\hline
\hline
Model  &  Parametrization  \\
\hline
BCFY~\cite{Braaten:1993jn,Braaten:1994bz}& $\frac{x(1-x)^{2}}{\left[1-(1-r)x\right]^{6}}\ \left[3+{\sum_{i=1}^{4} 
(-x)^{i}f_{i}(r)}\right]$ \\
  & \\
Bowler~\cite{Bowler:1981sb} & $x^{-(1+r_{b}b\ m_{\perp}^{2})} (1-x)^{a} \exp\left(-bm_{\perp}^{2}/x\right)$
       \\
  & \\
Colangelo--Nason~\cite{Colangelo:1992kh} &$ x^b(1-x)^a$ \\ & \\
Collins--Spiller~\cite{Collins:1984ms}&$\frac{\left( (1-x)^2+x(2-x)\epsilon_{b} \right) \left(1+x^2\right)}{x(1-x)\left(1-x^{-1}-\epsilon_{b}(1-x)^{-1}\right)^2}$\\  & \\
Kartvelishvili~\cite{Kartvelishvili:1977pi} & $x^{\alpha_{b}}(1-x)$  \\
 & \\
Lund~\cite{Andersson:1983ia}& $x^{-1}\left(1-x\right)^{a} \exp\left(-b\ m_{\perp}^{2}/x\right)$
      \\
 & \\
Peterson {\it et al.}~\cite{Peterson:1982ak} & $x^{-1}\left(1-x^{-1}-\epsilon_{b}\left(1-x\right)^{-1}\right)^{-2}$   \\
\hline
\hline
\end{tabular}
\end{center}
\end{table}}

Similarly to the measurements of light hadron production discussed in Section~\ref{sec:datalight}, the spectra of bottom mesons have been measured in $\epem$ collisions at the $Z^0$ pole by the LEP experiments~\cite{Heister:2001jg,Abreu:1995mu,Abbiendi:2002vt,Adeva:1991iw,Alexander:1995aj} and by the SLD collaboration~\cite{Abe:1999ki,Abe:2002iq} with an excellent statistical accuracy. In particular, both the OPAL and SLD data tend to exclude specific parametrizations of fragmentation functions such as Peterson {\it et al.}, Collins--Spiller and BCFY, favouring instead the Lund, Bowler, and to a lesser extent Kartvelishvili models~\cite{Abbiendi:2002vt,Abe:2002iq}. Bottom production has also been measured in DIS by the H1 collaboration~\cite{Adloff:1999nr} as well as in hadronic collisions  by UA1~\cite{Albajar:1990zu} and more recently by CDF~\cite{Abe:1993hr} and D\O~\cite{Abbott:1999se} at the Tevatron.

The measurement of charmed mesons in $\epem$ collisions at LEP energy is slightly more delicate than bottom production since momentum spectra need to be corrected for important $B$-meson weak decays. Still, rather precise measurements have been performed at LEP~\cite{Akers:1994jc}. However the bulk of data regarding $D$-meson production has been taken at lower energy. For quite some time, the only data available at $\sqrts=10$~GeV by the ARGUS~\cite{Albrecht:1991ss} and CLEO~\cite{Bortoletto:1988kw} collaborations had large statistical uncertainty and rather poor momentum resolution. Since then, high-precision measurements of the $D^0$, $D^+$, $D^{\star+}$, and $D^{\star0}$ spectra have been performed by BaBar~\cite{Aubert:2002ue},  BELLE~\cite{Seuster:2005tr} and CLEO~\cite{Artuso:2004pj}. Finally, charm measurements were also carried out in DIS by H1~\cite{Aktas:2004ka} and ZEUS~\cite{Breitweg:1999ad}, yet with a slightly lesser accuracy. Interestingly, the $D^{\star\pm}$ momentum spectra from ZEUS also disfavour the Peterson non-perturbative fragmentation functions~\cite{Breitweg:1999ad}.

The importance to get stringent constraints from $\epem$ data on the non-perturbative fragmentation function (and in particular its first moments) to predict charmed and bottomed meson production in hadronic collisions has been highlighted recently in $p$--$\bar{p}$ collisions at the Tevatron. The CDF collaboration reported that their measurement of $B^+$ spectra was a factor of 3 in excess over the ``QCD background''~\cite{Acosta:2001rz}. However, as shown in~\cite{Cacciari:2002pa}, this conclusion was actually drawn on the basis of a QCD calculation using the Peterson {\it et al.} FF that is excluded from LEP data. Predictions using more accurate non-perturbative fragmentation functions which reproduce well the first $N\le 10$ moments of the $B$-meson spectra at LEP, somewhat increase the bottom cross section thereby reducing the excess of CDF data\footnote{The final CDF measurements also slightly decreased as compared to the first preliminary results.} to a factor 1.7. Such an enhancement could then easily be explained by the present uncertainties on parton densities and the scale dependence of the FONLL calculation~\cite{Cacciari:2002pa}.

\begin{figure}[htb]
    \begin{center}
      \includegraphics[width=8.6cm]{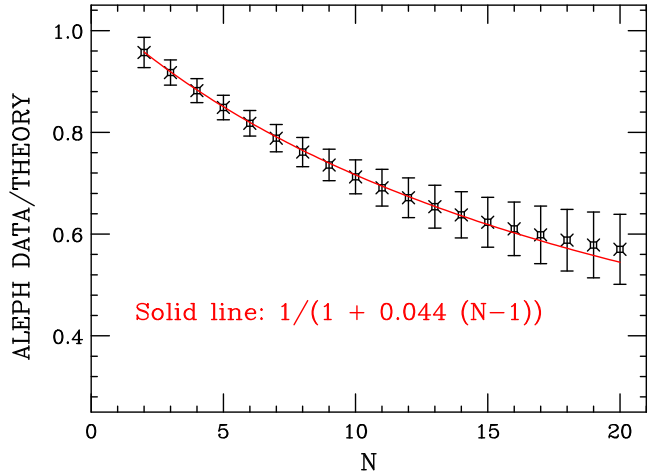}
    \end{center}
    \caption{ALEPH $D^{\star+}$ data compared to QCD predictions in momentum space, fitted by a simple functional form $1/(1+0.044(N-1))$. Taken from Ref.~\cite{Cacciari:2005uk}.}
    \label{fig:powercorrections}
\end{figure}

In order to illustrate the phenomenology of heavy-flavoured production, we briefly summarize in what follows the main results of a recent and detailed analysis of $D$ and $B$-meson production in $\epem$ collisions and the B-factories and LEP energy~\cite{Cacciari:2005uk}.
The analysis is performed using the NLO initial conditions (\ref{eq:dQQ})--(\ref{eq:dqlightQ}) for heavy-quark FF together with NLO coefficient functions $C_i$, both terms being properly resummed to NLL accuracy to account for soft gluon emission. The  NLO evolution of fragmentation functions is carried out with a proper matching conditions when crossing the bottom threshold~\cite{Cacciari:2005ry}, but neglecting any gluon splittings, $g\to Q\bar{Q}$, which are pretty much suppressed as already mentioned. Finally, the non-perturbative fragmentation function is taken from an average of a singular contribution, $\propto\delta(1-x)$ and the Colangelo--Nason parametrization~\cite{Colangelo:1992kh} (see Table~\ref{tab:npff}). Within this approach (and taking into account for possible initial-state radiation in $\epem$ collisions), the spectra of $D^{\star+}$, $D^{\star0}$ and $D^{\star+}\to D^0/D^{\star0}$ prove in excellent agreement with BELLE~\cite{Seuster:2005tr} and CLEO~\cite{Artuso:2004pj} data at $\sqrts\simeq10.6$~GeV. Interestingly, however, the extrapolation of these calculations to LEP energies turns out to be in contradiction with ALEPH measurements (which are the most precise so far) at large $x$. The disagreement between theory and the ALEPH data is illustrated in Figure~\ref{fig:powercorrections} where the data/theory ratio is plotted as a function of the moment of the fragmentation functions $F^H$. The reason for this disagreement may come from large power corrections to the $\epem$ coefficient functions, $C_1/\sqrts$ or $C_2/s$, whose strength can be fitted to ALEPH data, $C_1=0.57\pm0.16$~GeV or a somewhat larger value $C_2=5.1\pm0.3$~GeV$^2$. Unfortunately the lack of data at intermediate scales between the $\Upsilon(4S)$ and the $Z^0$ mass does not allow for a clarification of the scale-dependence of such corrections.

\section{Medium modifications}\label{sec:ffmed}

After having reviewed the experimental and theoretical studies of  fragmentation functions, we address in the second part of the paper the possible modifications of fragmentation processes induced by the presence of dense QCD media, such as nuclear matter in electron-nucleus scattering or quark-gluon plasma (QGP) in heavy-ion collisions.

\subsection{Data}\label{sec:datamedium}

As we have seen in Section~\ref{sec:ffvac}, the wealth of data in $\epem$, DIS and hadronic collisions allow for detailed studies of fragmentation processes ``in the vacuum''. Unfortunately, the clean $\epem$ process is obviously of no use in order to investigate possible medium modifications to fragmentation functions. Therefore, one has to rely either on DIS experiments on nuclear targets, $e$--A, or heavy-ion collisions, A--A; the former being
 ideal to probe cold nuclear matter\footnote{Proton-nucleus scattering is also interesting to probe fragmentation in nuclear matter, yet it is not as clean as DIS on nuclei.}  while the latter is devoted to the study of hot and dense --~though expanding~-- QCD medium. Data samples involving nuclei in the initial state are however much more scarce and kinematically limited than those involving protons.

\subsubsection{DIS on nuclear targets}

\begin{figure}[htb]
    \begin{center}
      \includegraphics[width=8.6cm]{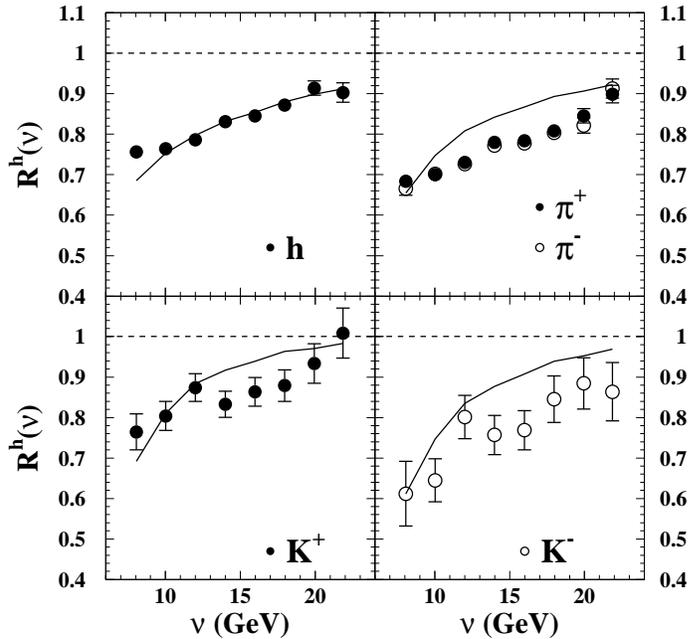}
    \end{center}
    \caption{Quenching of charged hadron, pions and kaon spectra in Kr targets measured by HERMES~\cite{Airapetian:2003mi} and compared to energy loss calculations~\cite{Arleo:2002kh,Arleo:2003jz}. Taken from Ref.~\cite{Arleo:2002kh}.}
    \label{fig:hermes}
\end{figure}

The HERMES experiment at DESY has performed extensive measurements of hadron production in semi-inclusive DIS (SIDIS) on various nuclear targets: N, Ne, Cu, Kr and Xe~\cite{Airapetian:2000ks,Airapetian:2007vu,Airapetian:2003mi}. Further, HERMES was able to perform the identification of $\pi^+$, $\pi^-$, $\pi^0$, $K^+$, $K^-$, $p$, and $\bar{p}$ {\it separately}. The hadron yield in a given ($\nu, x$)-bin, where $\nu$ is the virtual photon energy in the target rest frame and $x=E_h/\nu$ the hadron energy fraction, is normalized to the number of DIS inclusive events, $N_{\A}^e$. Assuming collinear factorization to be valid for hadron production in SIDIS on nuclei, the leading-order ratio can be written as:
\begin{equation}\label{eq:multDIS}
\frac{1}{N_{\A}^e}\frac{\dd N_{\A}^h(\nu,x)}{\dd\nu\,\dd{x}}=\left[\int\dd\xbj\sum_{i=q,\,\bar{q}}\sigma^{\gamma^* q}  f_i^{A}(\xbj, Q^2) D_i^{h/A}(x, Q^2)\right]
 \biggr/  \left[\int \dd\xbj \sum_{i=q,\,\bar{q}}  \sigma^{\gamma^* q}  f_i^{A}(\xbj, Q^2)\right],
\end{equation}
where $Q$ is the virtuality of the photon and $\sigma^{\gamma^* q}=\sigma^{\gamma^* q}(\xbj,\nu)$ the LO $\gamma^* q$ cross section. The fragmentation function $D_i^{h/A}$ indicates that the hadron is produced off a nuclear target, $e \A\to e\ h\ \X$. When $x$ is not too small, which is the case of the HERMES experiment, $x\gtrsim0.02$, sea-quarks in the target are small and therefore only up valence quarks contribute to the cross section (the contribution from down quarks being suppressed because of their smaller electric charge squared, $e_d^2=e_u^2/4$). In this approximation, the ratio simply reduces to the up-quark fragmentation function:
\begin{equation}\label{eq:multDISapprox}
\frac{1}{N_{\A}^e}\frac{\dd N_{\A}^h(\nu,x)}{\dd\nu\,\dd{x}}\simeq D_u^{h/A}(x, Q^2).
\end{equation}
It is indeed particularly interesting since the large current uncertainties on the nuclear parton densities, $f_i^{\A}$,  appearing in (\ref{eq:multDIS}) cancel out in this ratio\footnote{The first measurements of hadron attenuation in DIS on nuclei is due to the EMC collaboration~\cite{Ashman:1991cx}. Since hadron production was {\it not} normalized to the number of DIS events, the reported suppression in $e$--A with respect to $e$--$p$  is actually the consequence of the modifications of parton densities, the so-called \dots ``EMC effect''.}. The double production ratio in a heavy nucleus over that in a deuterium target~\footnote{We write $D_u^{h/D}=D_u^h$ in (\ref{eq:suppDIS}) since we do not expect any modification of fragmentation in such a small nucleus.},
\begin{equation}\label{eq:suppDIS}
R_A^{h}(x,\nu) = \frac{1}{N_{\A}^e}\,\frac{\dd N_{\A}^h(x,\nu)}{\dd\nu\,\dd x}\bigg/\frac{1}{N_D^e}\,\frac{\dd N_D^h(x,\nu)}{\dd\nu\,\dd x}\simeq D_u^{h/A}(x, Q^2) \bigg/ D_u^h(x, Q^2),
\end{equation}
(with $Q^2=2m_N \xbj$, $m_N$ being the nucleon mass) is then determined as a function of $\nu$ or $x$. As an example, the quenching of charged hadron, pion and kaon spectra measured as a function of $\nu$ by HERMES~\cite{Airapetian:2003mi} are shown in Figure~\ref{fig:hermes}.

Those data therefore prove ideal in order to study the detailed kinematic dependence of the medium-modified fragmentation process. This is will be of course only true if we can attribute the deviation of $R_A^{h}(z,\nu)$ from unity to be due to a modification of the up quark fragmentation into the hadrons. In that sense, the factorization formula Eq.~(\ref{eq:multDIS}) has to be seen as a model assumption. Suppose for instance that the typical time to form a hadron is of the order or less than the nuclear size, then inelastic rescattering of that very hadron in the nuclear medium may be responsible for the relative suppression of hadron production in $e$--A scattering, i.e. $R_A^{h}(z,\nu)<1$. Various models based on this ``hadron absorption'' mechanism have been proposed to account for the trend of the data~\cite{Falter:2003di,Accardi:2002tv,Accardi:2007in}. 

The CLAS experiment at JLab is currently performing similar measurements~\cite{Hafidi:2006ig}, yet presently at somewhat smaller energies, $\nu=2$--$3$~GeV, $Q^2\sim1$~GeV$^2$, for which perturbation theory may no longer be under control. Hopefully the upgrade at 12 GeV will allow for a more systematic study at higher energies ($\nu=2$--$9$~GeV, $Q^2=2$--$9$~GeV$^2$) of all hadron species~\cite{Hafidi:2006ig}, including all light mesons and strange baryons,   shedding new light on medium-modified fragmentation in cold nuclear matter.

\subsubsection{Heavy-ion collisions}

Large-$\pt$ hadron spectra have been measured in Au--Au collisions at RHIC ($\sqrtsnn=200$~GeV) by the BRAHMS, PHENIX, PHOBOS and STAR experiments. What is often reported experimentally is the ratio of the hadron yield at a given $\pt$ in Au--Au collisions (either minimum bias or in a centrality class ${\cal C}$) over that in $p$--$p$ collisions, properly scaled by the number of nucleon-nucleon (NN) binary collisions in that centrality class:
\begin{equation}\label{eq:quenchingfactor}
R^h_{_{\cal C}}(p_{_\perp}) = \frac{1}{\langle N_{{\rm coll}} \rangle \big |_{_{\cal C}}}\,\times\, \frac{\dd{\cal N}_{_{\rm {Au~Au}}}^{h}}{\dd{p_\perp} \dd y} \biggr/  \frac{{\dd{\cal N}_{_{p p}}^{h}}}{\dd{p_\perp} \dd y}.
\end{equation}
The number $\langle N_{{\rm coll}} \rangle \big |_{_{\cal C}}$ is determined from the Glauber multiple scattering theory~(see e.g. Appendix~I of~\cite{Arleo:2003gn}). It is expected to be $\langle N_{{\rm coll}} \rangle \big |_{_{{\cal C}\le20\%}}=779$ for the $20\%$ most central Au--Au collisions at RHIC energy. Since hard processes, such as large-$\pt$ production, are expected to scale with $\langle N_{{\rm coll}} \rangle$, the quenching factor (\ref{eq:quenchingfactor}) is $R^h_{_{\cal C}}(p_{_\perp})=1$ in the absence of any ``nuclear'' effect, which may come either from the properties of the initial state, e.g. modifications of parton densities in the incoming nuclei, either from final state interaction in a dense QCD medium.

\begin{figure}[htb]
    \begin{center}
      \includegraphics[width=9.6cm]{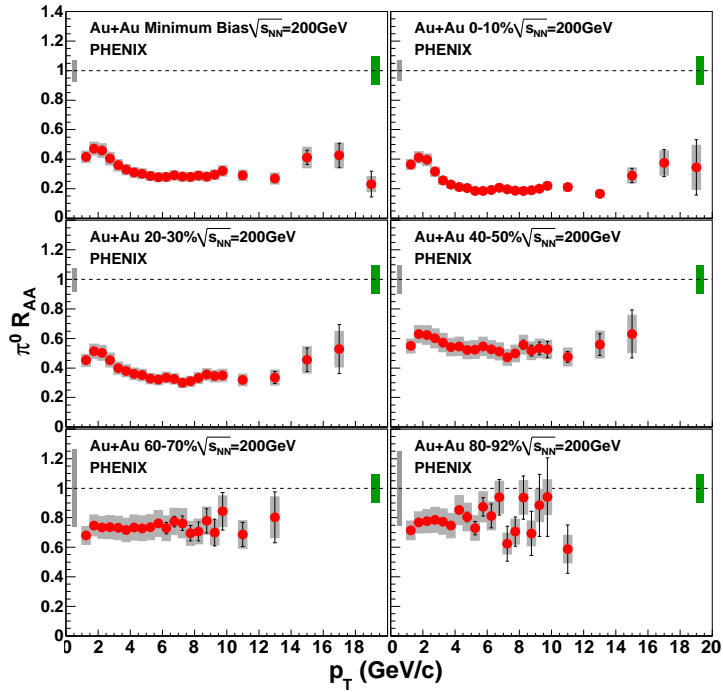}
    \end{center}
    \caption{Quenching factor of neutral pion production in Au--Au collisions at $\sqrtsnn=200$~GeV for various centralities measured by PHENIX. Taken from Ref.~\cite{Adare:2008qa}.}
    \label{fig:phenix}
\end{figure}

In central Au--Au collisions at $\sqrtsnn=200$~GeV, a significant suppression of neutral pions, $R^{\pi^0}\simeq0.2$ up to $\pt\simeq 20$~GeV has been reported by PHENIX and PHOBOS~\cite{Roland:2005yw}, a similar attenuation for $\eta$-mesons up to $\pt\simeq10$~GeV being also reported by PHENIX~\cite{Adler:2006hu}. STAR~\cite{Adams:2003kv} (and also BRAHMS~\cite{Arsene:2003yk} and PHENIX~\cite{Adler:2003au}) observes a similar quenching of the inclusive charged hadron yields ($\pt\lesssim10$~GeV). High-$\pt$ baryons ($p$, $\bar{p}$, $\Lambda$, $\bar{\Lambda}$) appear also suppressed, but at $\pt$ values higher than for mesons~\cite{Abelev:2007ra}. Interestingly, the hadron suppression pattern is less pronounced when going to more peripheral collisions~\cite{Adams:2003kv}; in particular, $R^h\lesssim1$ for the most peripheral (${\cal C}=80$--$92\%$) collisions, see Figure~\ref{fig:phenix}. Similar features are also observed at $\sqrts=130$ GeV\cite{Adcox:2001jp,Adcox:2002pe,Adler:2002xw}. For a complete discussion of RHIC measurements in heavy-ion collisions, we refer you to the comprehensive reviews by BRAHMS \cite{Arsene:2004fa}, PHENIX~\cite{Adcox:2004mh} and STAR~\cite{Adams:2005dq}, or to Ref.~\cite{dEnterria:2004nv} for a more concise survey of large $\pt$ measurements at RHIC.

These experimental results, sometimes referred to as ``jet quenching'' [{\it sic}], certainly one of the most spectacular phenomenon observed at RHIC, is often quoted as one big piece of evidence for parton energy loss in QGP discussed in the next Sections. What is particularly remarkable indeed is the absence of suppression in $d$--Au collisions where the quenching factor is around unity~\cite{Adams:2003im}. This is the indication that the mechanism responsible for the pion quenching is due to final state interactions --~hence for which parton energy loss is a natural candidate~-- and cannot be attributed to e.g. strong modifications of the parton densities in the colliding nuclei.

What happens then at lower energy? The limited phase-space available $(\sqrtsnn\simeq20$~GeV) at the SPS fixed-target facility does not allow for a systematic study of ``large'' $\pt$ hadron spectra. The WA98 $\pi^0$ measurements indicated an enhancement of the yield in Pb--Pb central collisions~\cite{Aggarwal:2001gn}. However, a re-analysis of the $\pi^0$ reference spectra in $p$--$p$ scattering leads to a quenching factor $R^{\pi^0}\simeq1$ or even below 1 in the most central collisions~\cite{dEnterria:2004ig}. Hopefully the data at RHIC intermediate energies will clarify this issue~\cite{Adare:2008cx}.

To study parton energy loss and medium-modified fragmentation functions would ideally require the reconstruction of jets in heavy-ion collisions; the huge background makes of course this task highly delicate. Nevertheless, thanks in particular to important theoretical developments on the jet reconstruction algorithms in a high-multiplicity environment~\cite{Cacciari:2005hq}, preliminary measurements by STAR~\cite{Putschke:2008wn}, but also studies at the LHC by ALICE~\cite{Alessandro:2006yt}, ATLAS~\cite{Grau:2008ef}, and CMS~\cite{dEnterria:2008ge}, look very promising.

\subsection{Parton energy loss}\label{sec:energyloss}

\subsubsection{Induced gluon spectrum}

Many signatures for quark-gluon plasma (QGP) formation have been suggested. Among them, it was argued first by Bjorken in the early eighties~\cite{Bjorken:1982tu} that hard quarks and gluons propagating through QGP may experience multiple scattering and therefore lose some energy because of the induced gluon emission. Experimentally, the immediate consequence would be a depletion of high-energy jets in heavy-ion collisions with respect to $p$--$p$ scattering. 

This process was then revived a decade later when Thoma and Gyulassy~\cite{Gyulassy:1992xb} and Gyulassy, Pl\"umer, and Wang ~\cite{Wang:1995fx} computed perturbatively the radiative energy loss of high-energy partons in QGP. Since then various approaches have been developed to determine the gluon radiation spectrum, $\dd I/\dd\omega$,  of hard partons undergoing multiple scattering. 

Baier, Dokshitzer, Mueller, Peign\'e, and Schiff (BDMPS) developed a perturbative framework to describe the medium-induced gluon emission process from the soft multiple scattering of hard partons in cold~\cite{Baier:1997sk} and hot~\cite{Baier:1997kr} QCD matter. The calculation assumes that the number of collisions, or opacity, is large: $n = L/\lambda\gg 1$, where $L$ is the medium length and $\lambda$ the parton mean free path. At small energy $\omega$, the typical time to produce gluons exceeds its mean free path, $t\sim\omega^{-1}\gtrsim\lambda$. Consequently several scattering centres act coherently to stimulate gluon emission, giving rise to the Landau-Pomeranchuk-Migdal (LPM) effect in QCD. Because of the destructive interference, the LPM spectrum is suppressed in the infrared, $\omega\dd{I}/\dd\omega\propto\omega^{-1/2}$, as compared to the independent Bethe-Heitler gluon spectrum, $\omega \dd{I}/\dd\omega\propto\omega^{-1}$. The BDMPS approach is equivalent~\cite{Baier:1998kq} to the powerful path-integral formulation of Zakharov~\cite{Zakharov:1997uu}, later generalized by Wiedemann~\cite{Wiedemann:2000za} for an arbitrary number of scatterings. Another approach by Gyulassy, L\'evai and Vitev has been carried out for thin media, $n=1,2\dots$, from which the gluon spectrum at any opacity is obtained recursively~\cite{Gyulassy:1999zd}. Let us also mention the finite-temperature approach by Arnold, Moore and Yaffe (AMY)~\cite{Arnold:2002ja,Arnold:2001ms} for moderately ``hard'' partons with energy $\cO{T}$, which was first developed to address the thermal photon rate in hot QGP~\cite{Arnold:2001ms}. Finally, a twist expansion has been proposed in Ref.~\cite{Guo:2000nz}. Since the goal in this paper is to address mostly fragmentation processes and their modifications, we refer the reader to~Ref.~\cite{Baier:2000mf} for a more comprehensive discussion on the different energy loss formalisms.

\subsubsection{Quenching weights}

In order to model the effect of energy loss on parton fragmentation into hadrons, an essential tool is the probability distribution in the energy loss, or {\it quenching weight}. Neglecting interference effects among the radiated gluons, which are $\cO{\alphas}$, Baier {\it et al.}~\cite{Baier:2001yt} gives the quenching weight a simple Poisson expression:
\begin{equation}\label{eq:quenchingweight}
{\cal P}(\epsilon) = \sum^\infty_{n=0} \, \frac{1}{n!}\ \left[ \prod^n_{i=1} \, \int \, \dd\omega_i \, \frac{\dd{I}(\omega_i)}{\dd\omega} 
\right] \delta \left(\epsilon - \sum_{i=1}^n  \omega_i\right)\ \exp \left[ - \int \dd\omega\ \frac{\dd{I}}{\dd\omega} \right],
\end{equation} 
in which the gluon spectrum is the only ingredient entering the calculation. The quenching weight helps to make the connection between the parton energy loss in QCD on the one hand and the experimental consequences on the other hand. In the general case, ${\cal P}(\epsilon)$ has a discrete contribution on top of a continuous part $p(\epsilon)$~\cite{Salgado:2003gb},
\begin{equation}\label{eq:quenchingweight2}
{\cal P}(\epsilon) = p_0\ \delta(\epsilon) + p(\epsilon).
\end{equation} 
The probability for no-gluon emission is given by $p_0=\displaystyle\lim_{\omega\to0}\exp[-N(\omega)]$, where $N(\omega)$ is the number of gluons radiated with an energy larger than $\omega$,
\begin{equation*}
N(\omega) = \int_\omega^{+\infty} \dd\omega^\prime\ \frac{\dd{I(\omega^\prime)}}{\omega^\prime}.
\end{equation*} 
For asymptotically large media the BDMPS gluon multiplicity is infrared divergent, $N(\omega)\propto\omega^{-1/2}$, and therefore the probability for no energy loss vanishes. With proper kinematic bounds for the gluon emission in finite-length media, however, the spectrum becomes finite in the infrared and therefore the quenching weight acquires an explicit singular contribution, $p_0\ne0$ in (\ref{eq:quenchingweight2})~\cite{Salgado:2003gb}.

\begin{figure}[htb]
    \begin{center}
      \includegraphics[width=8.6cm]{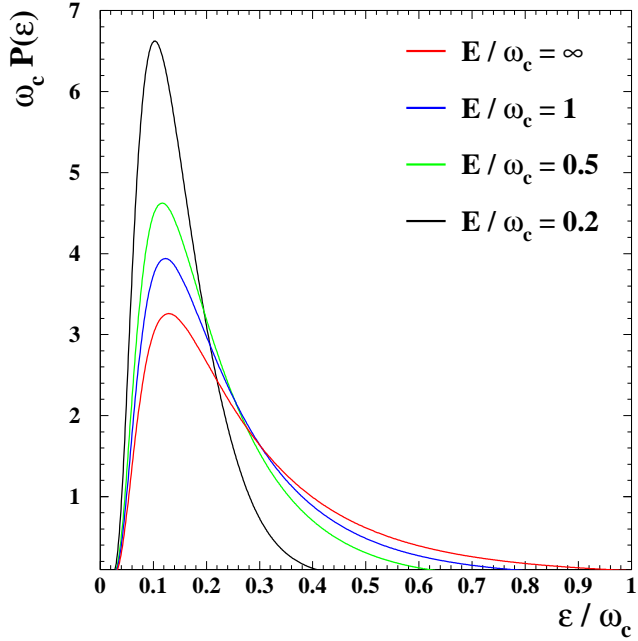}
    \end{center}
    \caption{Quenching weight using the BDMPS gluon spectrum for several quark energies in units of the energy-loss scale $\omega_c$. Taken from Ref.~\cite{Arleo:2002kh}.}
    \label{fig:quenchingweight}
\end{figure}

Using an integral representation of Eq.~(\ref{eq:quenchingweight}), the quenching weights have been computed using the BDMPS and Wiedemann induced gluon spectrum~\cite{Salgado:2003gb,Arleo:2002kh} and were made available for practical use either in terms of a numerical program~\cite{Salgado:2003gb} or an analytic parametrization~\cite{Arleo:2002kh}. For the sake of an illustration, the energy loss distribution is plotted in Figure~\ref{fig:quenchingweight} for various quark energies (in units of the energy loss scale, $\omega_c$).

Recently, Peshier~\cite{Peshier:2008bg} proposed a new way to determine the elastic, or collisional, quenching weight based on~(\ref{eq:quenchingweight}). As a first order Markov process, the momentum distribution of partons produced at a time $t+\delta t$ is given by that at time $t$,
\begin{equation*}
f_q(t+\delta t)={\cal T}_{qp}\ f_p(t)
\end{equation*}
in discrete momentum space, where the transition matrix ${\cal T}_{qp}$ is simply related to the $p\to{q}$ transition rate. Therefore the probability for a hard parton with energy $E(p)$  to lose an energy $\varepsilon=E(p)-E(q)$ in the time interval $\delta t$ is basically given by ${\cal T}_{qp}$ which, by successive iteration (i.e. matrix multiplications), gives the full quenching weight. On top of being numerically fast, one of the advantages of that method is that it should be appropriate for any parton energy, while the Poisson approximation (\ref{eq:quenchingweight}) explicit breaks down when considering finite parton energies as discussed in~\cite{Arleo:2002kh}.

\subsubsection{Expanding media}

The medium-induced gluon spectrum, and therefore the quenching weight, is determined for 
static and uniform media. However, the dense QCD medium produced in the early times of a heavy-ion collision experiences fast (and mostly longitudinal) expansion. In order to account for this dynamical expansion, the transport coefficient characterizing the scattering power of the medium, $\hat{q}$, is usually rescaled according to \cite{Salgado:2002cd},
\begin{equation}\label{eq:dynamicalscaling}
  \hat{q}(L) = \frac{2}{L^2}\int_{\tau_0}^{L}
  \dd\tau\, \left(\tau - \tau_0\right) \,
  \hat{q}(\tau),
\end{equation}
where $\alpha$ characterizes the time-dependence of the medium
energy-density, $n(\tau)\propto \tau^{-\alpha}$, and $\qhat(\tau) = \qhat(\tau_0) \left(\tau_0/\tau\right)^\alpha$. The purely longitudinal (or Bjorken) expansion corresponds to
$\alpha=1$, and is often assumed in phenomenological applications. Recently, a remarkably simple prescription has been given recently by Arnold in order to determine $\dd I/\dd\omega$ in a finite and expanding medium~\cite{Arnold:2008iy}. Therefore, applying this recipe using e.g. hydrodynamical space-time evolution will allow eventually for the computation of more realistic quenching weights without relying on the assumption~(\ref{eq:dynamicalscaling}).

\subsection{Kinematic rescaling}\label{sec:rescaling}

\begin{figure}[htb]
    \begin{center}
      \includegraphics[width=8.6cm]{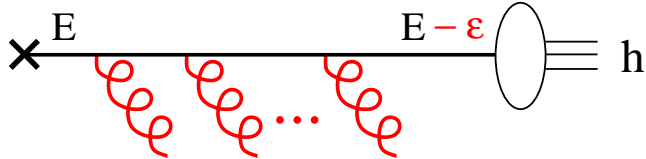}
    \end{center}
    \caption{Schematic view of the energy loss process and the medium-modified fragmentation function in the energy rescaling model.}
    \label{fig:modelFF}
\end{figure}

The first approach to model medium-modified fragmentation functions accounting for parton energy loss in QCD media is due to Wang, Huang and Sarcevic~\cite{Wang:1996yh}. The model basically assumes that the $Q^2$ evolution of the fragmentation process is not affected by the medium. The effect of energy loss is simply to rescale the initial parton energy, from say $E$ to $E-\epsilon$, after which the parton fragments as if in the vacuum (see sketch of the model in Figure~\ref{fig:modelFF}). Therefore, the medium-modified FF, call it ${\cal D}$, can be written as~\cite{Wang:1996yh}: 
\begin{equation}\label{eq:modelff}
 x {\cal D}_i^h(x, Q^2) = \int_0^{(1-x)E_i} \dd\epsilon \,{\cal P}(\epsilon)\, 
    x^\star\,D_i^h (x^\star, Q^2) + \int_{x E_i}^{E_i} \dd\epsilon\, {\cal P}(\epsilon)\ \frac{x E_i}{\epsilon}\ D_i^h\left(\frac{x E_i}{\epsilon}, Q^2\right),
\end{equation}
where the energy-fraction carried away by the hadron is shifted from $x$ to $x^\star\equiv x/(1-\epsilon/E_i)$. Because FF are steeply falling functions of $x$ (see Section~\ref{sec:ffvac}), even a small shift in the momentum fraction could lead to a significant hadron yield suppression. The second term in the r.h.s. of Eq.~(\ref{eq:modelff}) accounts for the fragmentation of the radiated gluons induced by the medium. Although it satisfies the momentum sum rule (\ref{eq:sumrule}), it is legitimate to question the use of the large-$x$ FF and at very low $Q^2$ in order to fragment those gluons into hadrons (a simpler LPHD prescription may appear more appropriate). Further, a hard parton which has lost an energy $\epsilon$ through the medium is likely to have emitted {\it several} gluons ($n\gg1$) whose energies sum up to $\epsilon$, $\displaystyle\sum_{i=1,n}\omega_i=\epsilon$, rather than a {\it single} gluon with energy $\omega_1=\epsilon$ (see e.g.~\cite{Arleo:2002kh}).

This model --~or variants of it~-- has been widely used to describe the hadron quenching measured at RHIC~\cite{Gyulassy:2001nm,Eskola:2004cr}. The trend of the data as a function of $\pt$ was in particular nicely reproduced when geometrical effects --~partons being produced anywhere in the dense medium~-- are taken into account~\cite{Eskola:2004cr}. The calculation based on the BDMPS/Wiedemann framework, however, led to values of the transport coefficient somewhat larger~\cite{Eskola:2004cr} than what is expected in perturbation theory~\cite{Baier:1997sk,Baier:2002tc}. More recently, progress has also been made towards the implementation of a full hydrodynamical expansion of the produced medium --~constrained by soft/global observables~-- in energy-loss scenarios, first in~\cite{Hirano:2002sc} and more recently in~\cite{Renk:2006sx} aiming at a consistent description of soft and hard probes in heavy-ion collisions.

In order to illustrate the qualitative features of the model, let us take the case of SIDIS reaction for simplicity. Forgetting about the convolution for the sake of the discussion, assuming ${\cal P}(\epsilon)=\delta(\epsilon-\langle\epsilon\rangle)$ as long as $x$ is not too large, the suppression of hadrons using (\ref{eq:modelff}) in (\ref{eq:suppDIS}) becomes,
\begin{equation*}
R_A^h(x, \nu) \simeq 1 + \frac{x\ \meaneps}{\nu}\ \frac{\partial\ln D_u^h(x)}{\partial x}.
\end{equation*}
First of all, the suppression is an increasing function of the photon, hence of the up quark energy in the nuclear rest frame. In particular, note that in the Bjorken limit $\nu \gg \meaneps$, hence $Q \gg \meaneps$ at finite $\xbj$, the effects of the final state interactions become negligible, ${\cal D}_i^h(x, Q^2)\simeq D_i^h(x, Q^2)$. Secondly, it is is interesting to notice that the suppression of hadrons is sensitive to the (logarithmic) slope of the ordinary fragmentation functions~\cite{Arleo:2003jz}. Hence, since the unfavoured fragmentation functions are {\it softer} than the favoured ones (see discussion in Section~\ref{sec:fitassumptions}), this in turn leads to a {\it larger} suppression of hadrons which do not contain up valence quarks (as long as $\xbj$ is not too small). This effect is actually seen in the HERMES data (see Figure~\ref{fig:hermes}) which report a stronger suppression of $K^-$ than $K^+$, $R_A^{K^-}<R_A^{K^+}$, in agreement with the energy loss model calculations\footnote{The absorption models also predict a stronger suppression of $K^-$ than $K^+$ because of a larger inelastic rescattering cross section in nuclear matter. See e.g. the short discussion in~\cite{Arleo:2003yf}.}~\cite{Arleo:2003jz}. Finally, at large $x$ the phase-space to emit gluons with energy less than $(1-x)E_u$ becomes very much restricted. Therefore the suppression shall be all the more stronger as $x$ gets larger, as can be seen from the upper limit of the integral (\ref{eq:modelff}). In the $x\to1$ limit, in particular, the DIS quenching factor is given by the probability for no-gluon radiation in the nuclear medium,
\begin{equation*}
R_A^h(x\to1,\nu) = p_0.
\end{equation*}
The $x$ dependence from the rescaling model turns out to be as well in rather good agreement with the HERMES measurements~\cite{Arleo:2003jz}; see also Ref.~\cite{Accardi:2007in} for a more recent discussion.

\subsection{Parton evolution in QCD media}\label{sec:evolutionmedium}

The above-discussed rescaling model appears reasonable to describe the energy loss of the leading parton. However, the rescattering of the full parton shower in the medium is not properly taken into account in the Poisson approximation, Eq.~(\ref{eq:quenchingweight}), of the quenching weights. Moreover, the $Q^2$ dependence of the parton energy loss process is not taken into account in that model (the medium-modified ${\cal D}(x, Q^2)$~(\ref{eq:modelff}) actually satisfies the standard DGLAP evolution equations).

In order to address these issues, several approaches using the MLLA resummation scheme, encountered in Section~\ref{sec:mlla}, have been proposed recently~\cite{Borghini:2005em,Armesto:2007dt,Domdey:2008gp}. The common idea of these studies is to modify the evolution kernel, $K_i^j$ in Eq.~(\ref{eq:eveq}), in order to model the extra gluon radiation in the case of radiative energy loss~\cite{Borghini:2005em,Armesto:2007dt}, or the rescattering of the emitted partons in the medium~\cite{Domdey:2008gp}.

In the approach of Borghini and Wiedemann~\cite{Borghini:2005em}, the gluon radiation induced by the medium modifies the singular part of the QCD splitting functions; for example,
\begin{align}
  P^{\rm med}_{qq}(z) = C_F \left[ \frac{2(1+f_{\rm med})}{(1-z)_+} - (1+z) \right],
\end{align}
where $f_{\rm med}$ is a parameter controlling the ``amount'' of induced gluon radiation. In particular, taking $f_{\rm med}\to0$ brings back to the ordinary splitting functions. The more infrared singular behaviour of the induced gluon spectrum, with respect to that in the vacuum, justifies the choice of modifying only the $1/z$ part of the splitting functions. Shifting splitting functions in order to reproduce parton energy loss in QCD media was first advocated in~\cite{Guo:2000nz}.

\begin{figure}[htb]
  \centering
  \includegraphics[width=8.9cm,height=7.8cm]{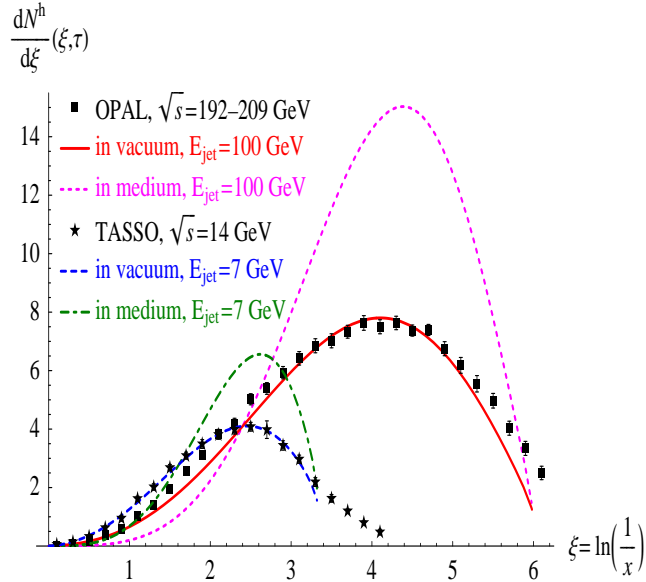} 
  \caption{Single inclusive hadron spectrum as a function of $\xi$ using vacuum splitting functions (compared to $\epem$ data) and using medium-modified splitting functions. Taken from Ref.~\cite{Borghini:2005em}.
  }
  \label{fig:distortinghbp}
\end{figure}

The medium-modified MLLA equation can therefore be solved in a very same way as Eq.~(\ref{eq:dglapsmallx}) in the vacuum in order to compute a variety of observables in presence of a dense QCD medium. The single inclusive spectrum $D^h(x, Q^2)$ has been determined in particular in~\cite{Borghini:2005em} and is reproduced in Figure~\ref{fig:distortinghbp}. As can be seen, the usual hump-backed plateau is ``distorted'' by the medium:  the number of highly-energetic particles (small $\xi$) is quenched as compared to the expected vacuum spectrum, while the soft gluon yield is dramatically enhanced at large $\xi$. This is nothing but the manifestation of the (approximate) momentum-conservation in the MLLA evolution: the medium converts part of the energy of a hard parton into the production of many soft partons (there is no ``energy loss'' as such!). This model was recently extended beyond the limiting spectrum approximation ($Q_0\ne\lqcd$), that is for hadrons with mass $m\simeq Q_0$~\cite{Sapeta:2007ad}. A specific dependence on the hadron species inside the jet (``jet hadrochemistry'') is predicted and could be tested with ALICE at LHC which has good identification capabilities~\cite{Alessandro:2006yt}, as well as CMS~\cite{dEnterria:2008ge} to a certain extent.

 In the approach of Armesto, Cunqueiro, Salgado and Xiang (ACSX)~\cite{Armesto:2007dt}, the medium-modified splitting functions are directly related to the medium-induced gluon spectrum, $\dd I^{\rm med}/\dd{z}\ \dd{Q^2}$, computed in the Wiedemann parton energy loss framework~\cite{Wiedemann:2000za},
 \begin{eqnarray}
  P(z) &=& P^\text{vac}(z) + \Delta P(z, Q^2),\nonumber\\
 \Delta P(z, Q^2) &=& \frac{2\pi Q^2}{\alphas}\ \frac{\dd I^{\rm med}}{\dd{z}\dd{Q^2}}.
 \end{eqnarray}
As compared to the Borghini--Wiedemann model, the medium modifications are now explicitly dependent of the parton virtuality through the induced gluon spectrum. No energy loss effects are expected at the time of hadronization which occurs on long time scales. Therefore, the FF at the initial scale $Q_0^2\sim2$--$4$~GeV$^2$ is given by that in the vacuum (specifically, the KKP set~\cite{Kniehl:2000fe} is assumed)~\cite{Armesto:2007dt}. A virtue of this approach is that at large virtualities, $Q\gg\lqcd$, the rescaling model (\ref{eq:modelff}) using the quenching weights (\ref{eq:quenchingweight}) is formally recovered~\cite{Armesto:2007dt}. 

\begin{figure}[htb]
  \centering
  \includegraphics[width=8.cm,height=8.6cm]{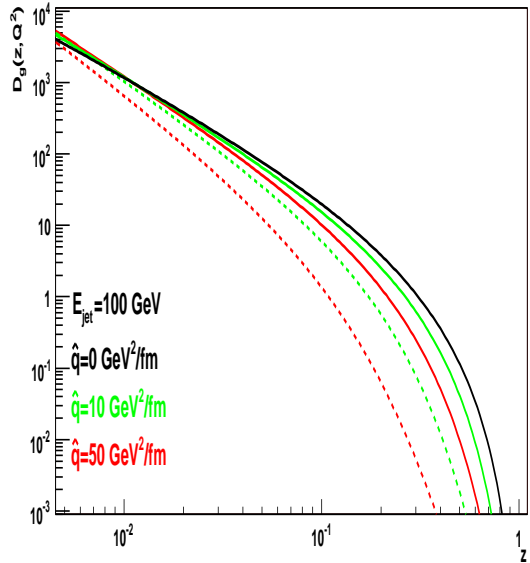} 
  \caption{Medium-modified gluon-to-pion fragmentation functions for different values of $\hat{q}$ and medium lengths. Taken from Ref.~\cite{Armesto:2007dt}.
  }
  \label{fig:acsx}
\end{figure}

The medium-modified FF have been determined on a large range of $x=5\times10^{-3}$--$1$ and for all $Q^2$~\cite{Armesto:2007dt}. As expected, the effects of parton energy loss is to soften the fragmentation functions, with a large suppression at high-$x$ and a clear enhancement at small $x$, see Figure~\ref{fig:acsx}. Of course, the larger the medium length or the transport coefficient, the stronger the medium effects. What is perhaps less intuitive is the $Q^2$ dependence: the effects of parton energy loss are more pronounced as $Q^2$ gets larger\footnote{Note that medium-modified FF do not coincide with those in the vacuum for asymptotically large $Q^2$ and at finite $x$ as one could have naively  thought for higher twist effects.}. This behaviour is explained in~\cite{Armesto:2007dt} by the longer time-evolution, even though the time spent by the hard parton in the medium may naively be set by the medium size and lifetime.  Within the same framework, the medium dependence of the mean and the dispersion of multiplicity distributions of partons inside quark and gluon jets has also been obtained analytically recently~\cite{Dremin:2006da,Quiroga:2008hp}.

The recent RHIC data on the suppression of single inclusive electron spectra in Au--Au collisions~\cite{Bielcik:2005wu} has triggered a renewed interest on elastic, or collisional, energy loss in QCD media~\cite{Peigne:2008nd}. In the same spirit as the above describe radiative energy loss models, it was suggested in~\cite{Domdey:2008gp} to take into account $2\to2$ scattering processes by the inclusion of a scattering function, ${\cal K}$, in the leading order evolution kernel. The function ${\cal K}$ is proportional to the medium gluon density and to the $gg\to{gg}$ scattering cross section,
\begin{equation*}
-\frac{\dd\sigma}{\dd{t}} = \alphas^2(Q^2)\ \frac94\ \frac{2\pi}{(-t+m_D^2)^2},
\end{equation*}
where the Debye mass, $m_D$, acts as an infrared regulator. A large suppression is predicted at high $x$ but, unlike the radiative energy loss approach~\cite{Armesto:2007dt},  no enhancement is expected at small values of $x$. For consistency, however, it would be interesting to include such a term, ${\cal K}=\cO{\alphas^2}$, in a DGLAP equation at NLO, and to check whether the medium effects remain similar.

Finally, let us also mention that fragmentation functions including parton multiple scattering have been determined in a higher-twist framework~\cite{Guo:2000nz}. Although the calculation was first applied in the context of electron-nucleus collisions with phenomenological success, it has also been extended for the case of hot QCD media. Recently, Aurenche, Zakharov, and Zaraket (AZZ)~\cite{Aurenche:2008hm} however showed that the correct gluon emission spectrum actually vanishes when the number of rescattering is $n=1$; this claim is disputed by Wang in~\cite{Wang:2008ti}, yet maintained by AZZ in~\cite{Aurenche:2008mq}.

\subsection{Medium-modified parton showers}\label{sec:partonshowers}

Parton showers (PS) based on Monte Carlo techniques, e.g. {\bf HERWIG}~\cite{Corcella:2002jc} or {\bf PYTHIA}~\cite{Sjostrand:2006za}, share many advantages. First of all, they allow for the energy-momentum conservation throughout the full evolution, unlike analytic approximations such as DLA or MLLA. Moreover, PS are able to characterize the underlying event activity in the collisions --~an important issue with heavy-ions~-- and to ease the comparison with experimental measurements. Finally, they also allow for a better understanding of the microscopic dynamics.

\begin{figure}[htb]
  \centering
  \includegraphics[width=12.4cm]{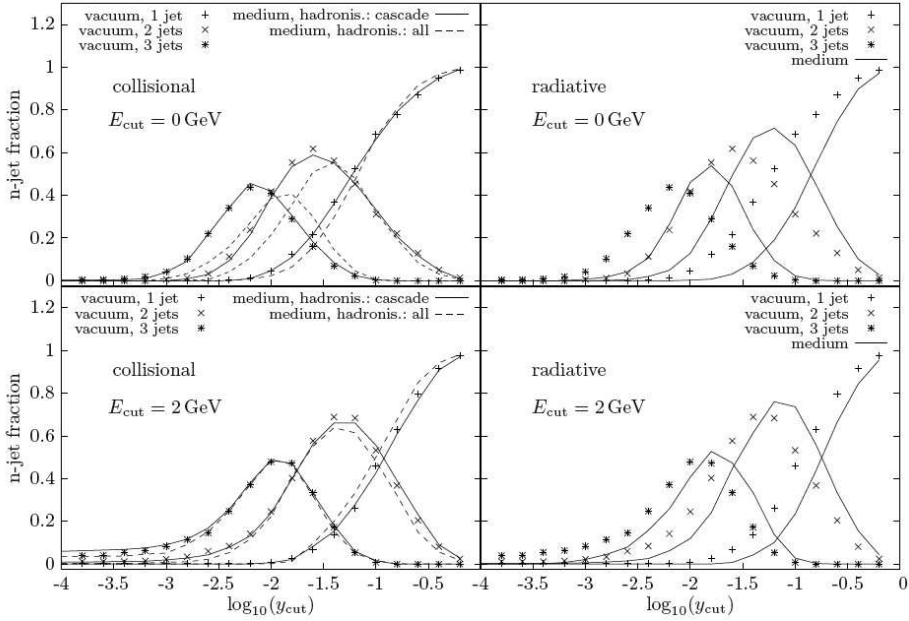} 
  \caption{$n$-jet distribution predicted by {\bf JEWEL} with elastic (left) and radiative (right) energy loss. Taken from Ref.~\cite{Zapp:2008gi}}
  \label{fig:jewel}
\end{figure}

In heavy-ion collisions, several parton shower models which describe the jet evolution in QCD media have been developed recently~\cite{Zapp:2008gi,Renk:2008pp}. In~\cite{Zapp:2008gi}, the Monte Carlo programme {\bf JEWEL} ({\bf J}et {\bf E}volution {\bf W}ith {\bf E}nergy {\bf L}oss) is first tested for various observables, such as jet shapes in $\epem$ collisions. Concerning the medium modifications, radiative energy loss is implemented using the model by Borghini and Wiedemann~\cite{Borghini:2005em} described in Section~\ref{sec:evolutionmedium} where $f_{\rm med}$ is a free parameter. Elastic rescattering is also included in {\bf JEWEL} through the aforementioned $2\to2$ scattering process~\cite{Domdey:2008gp}. Therefore, exploring various jet observables with {\bf JEWEL} may allow for both energy loss mechanisms to be disentangled. A good example is the $n$-jet fraction as a function of the resolution scale $\log_{10}(y_0)$ assumed in the jet algorithm. Figure~\ref{fig:jewel} shows for instance that elastic rescattering almost has no effects on the vacuum $n$-jet distribution, while on the contrary the number of jets produced at a given $y_0$ increases somehow when radiative energy loss are taken into account~\cite{Zapp:2008gi}. This is an encouraging step towards the understanding of the mechanisms responsible for jet quenching. Another PS~\cite{Renk:2008pp} was also developed from a modification of the {\bf PYTHIA} shower algorithm {\bf PYSHOW}~\cite{Bengtsson:1986hr}. The effect of parton energy loss is to increase the virtuality of the rescattering partons (thus making them to radiate more) by an amount $\Delta Q^2$ which directly depends on the space-time dependence of the transport coefficient $\hat{q}=\hat{q}({\bf x}, t)$ given by a 3D hydrodynamical evolution. Finally, note that the approach of ACSX~\cite{Armesto:2007dt} is being implemented in {\bf PYTHIA} as well and preliminary results have already been given~\cite{Cunqueiro:2008hp}.

\subsection{Measuring (medium-modified) fragmentation functions}\label{sec:correlations}

There is an obvious need to identify observables in heavy-ion collisions which may be sensitive to parton energy loss in hot QCD media. As stressed e.g. in~\cite{Eskola:2004cr}, the single inclusive hadron $\pt$-spectra is unfortunately not to sensitive to the microscopic dynamic underlying the ``jet quenching'' phenomenon. As mentioned in Section~\ref{sec:datamedium}, there is a real hope that measuring spectra of hadrons inside reconstructed jets, yet difficult, may be possible. In this Section another observable is explored: the double inclusive production of a prompt photon and a large-$\pt$ hadron.

Consider the leading-order QCD Compton process for instance, $qg\to{q}\gamma$. Because of momentum conservation, the photon is produced back-to-back to the hard quark, with equal and opposite transverse momentum. As a consequence, the $\gampi$ momentum imbalance variable,
\begin{equation*}
\z \equiv - \frac{{\bf \ptpi} . {\bf \ptgamma}}{|{\bf \ptgamma}|^2},
\end{equation*}
reduces to the fragmentation variable, $\z = x$, in this leading-order (LO) kinematics. Therefore, there should be in principle a clear connection between the {\it experimentally} accessible photon--hadron momentum-imbalance distributions on the one hand, and the {\it theoretical} quark fragmentation function into hadrons on the other hand, as first suggested in~\cite{Wang:1996yh}.

There a a few caveats though. First of all, the photon can itself be produced by the collinear fragmentation of a leading parton. In this case, there is no correlation between the hadron and the photon momenta since the two jets fragment independently. This is precisely the most important ``background'' channel which can get rid of using appropriate isolation criteria. Other limitations which could somehow complicate the picture are higher-order corrections as well as soft and collinear gluon radiation.

\begin{figure}[htb]
\centering
\includegraphics[width=8.6cm]{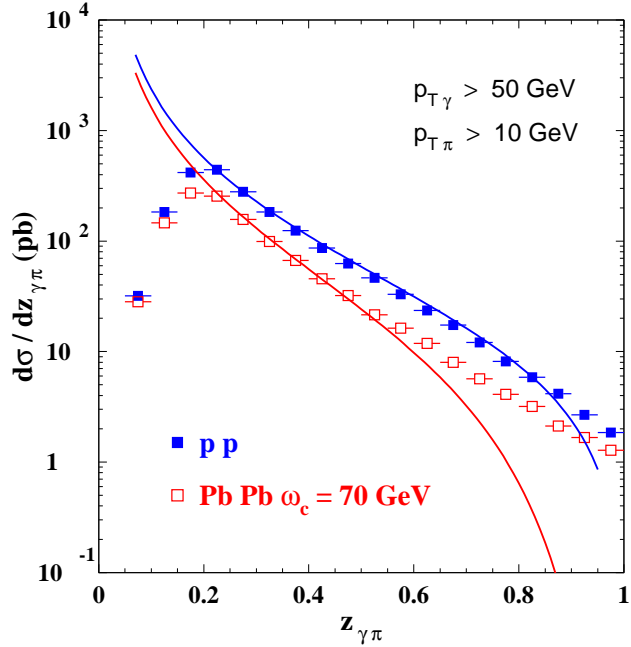}
    \caption{$\gampi$ imbalance distributions in $p$--$p$ and Pb--Pb collisions at the LHC (squares), as compared to the fragmentation functions used in the calculation (solid lines). Taken from Ref.~\cite{Arleo:2007qw}.}
\label{fig:distributionsgampi}
\end{figure}

In order to illustrate the interest of this observable, the $\gampi$ distributions computed in QCD at LO are shown in Figure~\ref{fig:distributionsgampi} for $p$--$p$ and Pb--Pb collisions at the LHC. The steeply falling shape above $\z\ge\picut/\gacut=0.2$ is reminiscent of that of parton fragmentation functions. In order to further show that similarity, the quark fragmentation functions into pions used in the calculation (either in the vacuum for $p$--$p$ collisions or modified by the medium in Pb--Pb) at virtuality $Q=\gacut$ are shown as solid lines. Note that at large $\z$, no matching between correlation distributions and fragmentation functions is observed because of the onset of the fragmentation-photon channel. For more details, we refer the reader to~\cite{Arleo:2004xj}.

\section{Outlook}

Parton fragmentation studies have been reviewed in this paper, trying to focus on the most recent advances of this active field.

In the vacuum, no doubt that the (still preliminary) data from the B-factories and the more precise measurements to be carried out at RHIC with higher luminosities will prove essential to constrain further the fragmentation functions into light and heavy hadrons, specifically in the baryon sector. More refined experimental tests of the fragmentation processes at small-$x$ shall also be achieved at the Tevatron; equally impressive results can be expected by the ATLAS and CMS experiment at the LHC. On the theory side, a renewed interest has emerged recently with a systematic investigation of corrections beyond MLLA and for a variety of jet observables (single spectra, multiplicity distributions, 2-particle correlations,\dots).

The understanding of QCD evolution in a medium is subject to a lot of attention. For years most of phenomenology has been based on the intuitive, yet rather simple, energy rescaling model. New interesting proposals have been suggested lately to go beyond that picture, with the modification of the evolution kernels in DGLAP or MLLA evolution equations. Despite those recent efforts, many questions are still open. It would be for instance interesting to investigate and understand how the QCD coherence effects observed for jets produced in the vacuum will manifest (or not) for jets traversing a dense and colourful medium; is there any angular ordering in QCD media?

Finally, despite the now very precise RHIC data, it becomes increasingly clear that more differential observables will be needed to constrain the medium-modifications of fragmentation functions. Hopefully the possible full jet reconstruction in heavy-ion collisions and the measurements of photon--jet processes shall soon be performed and shed new light on our understanding of parton energy loss and the formation of QGP in heavy-ion collisions.

\acknowledgement{It is a pleasure to thank the organizers of the Hard Probes 2008 conference, and in particular N\'estor Armesto and Carlos Salgado, for having set up this exciting event. I am also indebted to David d'Enterria, Redamy P\'erez Ramos y Paloma Quiroga Arias for useful comments on the manuscript.}

\providecommand{\href}[2]{#2}\begingroup\raggedright\endgroup

\end{document}